The Airbnb anti-discrimination team

# Measuring discrepancies in Airbnb guest acceptance rates using anonymized demographic data

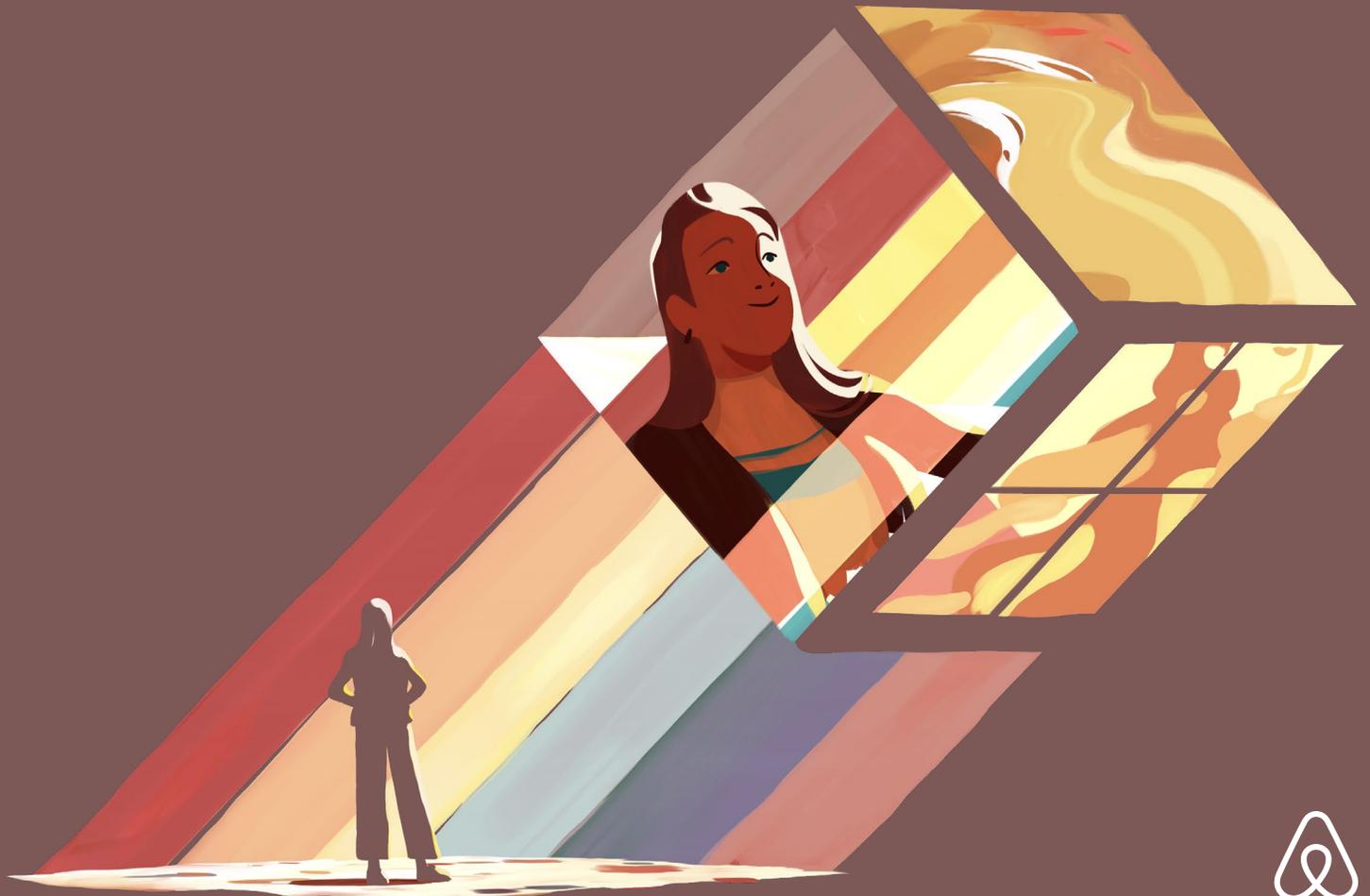
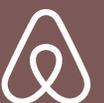

**The Airbnb anti-discrimination team is: Sid Basu\*, Ruthie Berman, Adam Bloomston\*, John Campbell, Anne Diaz\*, Nanako Era, Benjamin Evans, Sukhada Palkar, and Skyler Wharton.**

The symbol * denotes corresponding authors—they may be reached at antidiscrimination-papers@airbnb.com.

We would like to thank the following reviewers for their feedback: Dr. Latanya Sweeney and Jinyan Zang (Data Privacy Lab at Harvard); Dr. Cathy O'Neil and Jacob Appel (O'Neil Risk Consulting & Algorithmic Auditing); Antony Haynes, J.D. (Director of Cybersecurity and Privacy Law at Albany Law School); Gennie Gebhart, Kurt Opsahl, and Bennett Cyphers (Electronic Frontier Foundation); Harlan Yu, Aaron Rieke, and Natasha Duarte (Upturn); Mallory Knodel (Center for Democracy & Technology); and Dr. Conrad Miller (Assistant Professor of Economics at Berkeley Haas School of Business).

# Introduction

**Airbnb wants to create a world where anyone can belong anywhere.**

We take this mission to heart, so we were deeply troubled by stories of travelers who were turned away by Airbnb hosts during the booking process because of the color of their skin. #AirbnbWhileBlack became a trending hashtag, and we heard from many hosts and guests who shared painful stories of being discriminated against.

No online platform or marketplace is immune to the problems of racism and discrimination, including Airbnb. To help create a world where anyone can belong anywhere, we knew we had to do more to fight discrimination and bias.

We responded by making a series of commitments to combat discrimination [Murphy 2016]. In addition to an overhaul of our non-discrimination policy and improved responses to reports of discrimination, one of our more substantial commitments was to form a permanent anti-discrimination product team dedicated to fighting discrimination on Airbnb and ensuring that everyone has equal access to the trips they want to take. Over the past few years, the team has focused specifically on discrepancies in the product experience that may be attributable to race, and has introduced a number of product improvements intended to address the features of the online Airbnb experience that were adversely affecting some racial groups more than others. These discrepancies can be the result of explicit or implicit discrimination on an individual level (e.g., between hosts and guests), as well as the result of broader, systemic inequities [Edelman 2014, Edelman 2017]. Our most notable product change to date has been to time the display of the guest profile photos with a confirmed booking, in order to reduce bias in the booking request process.

In order to understand the true impact of such interventions, we must be able to measure how guests and hosts from different demographic groups experience Airbnb. Specifically, our team would like to measure *experience gaps* on Airbnb that are attributable to *perceived race*. We say that an *experience gap* occurs when one group of people is adversely impacted by a part of our product experience more than another group. This concept is best understood through a concrete example. Airbnb's homes booking process connects hosts, who list spaces where one can stay, to guests, who are searching for lodging. We can define a metric, the *acceptance rate*, as the rate at which hosts accept guests' reservation requests. We can then compute this metric for different groups of guests in our study. The gap between acceptance rates for guests in two different groups is an experience gap which, in this case, we call an *acceptance rate gap* (ARG).

Since discrimination is generally a result of one person's perception of another, we've chosen to concentrate on measuring inequities with respect to these perceptions, which we're calling *perceived race* [Smedley 2005]. There are many nuances in how these perceptions are made and influence human behavior that warrant further exploration. To narrow the focus of this initial paper, we concentrate on how to measure the acceptance rate gap between guests who are perceived to be white and guests who are perceived to be black within the United States.

It is important that we strive to provide the highest degree of privacy protection we can for Airbnb hosts and guests while we conduct this measurement. The problem of balancing data utility and privacy is a complex one, with no clear cut solution. Nevertheless, we believe that it is critically important to protect the privacy of our community. The risk of *misuse* of perceived race data is serious, whether from internal or external threats.

We have developed our system with the help, guidance, and insight of several civil rights partners, privacy groups, and academic advisors. The system is designed to minimize the risk that individuals' perceived race can be retrieved from the data that are used to make measurements. We achieve this by enforcing the privacy model of p-sensitive k-anonymity in order to prevent certain attribute disclosure; we also utilize asymmetric encryption across trust boundaries to ensure that data flows are one-way. Because anonymization can adversely affect data utility, we have developed a simulation-based methodology to understand the effect of data anonymization on statistical power. Our analysis shows that, even with anonymized[1] data, we can still achieve an adequate degree of statistical power to measure the efficacy of interventions to reduce the acceptance rate gap.

The impact that technology companies have on our daily lives cannot be overstated, and with that impact comes a tremendous responsibility to treat the people who rely on our products with integrity and respect. Our hope is that this technical paper can serve as a blueprint for other companies. We also hope it can be a starting point for further innovations that not only rigorously measure discrepancies present on online platforms, attributable to perceived demographics, but also uphold user privacy.

---

1 We define *anonymized*, *anonymity*, and *anonymization* as used throughout this technical paper in the section **Technical overview: Disclosure threat categories**.

# Technical overview

The goal of this paper is to present a system that analyzes the acceptance rate gap using anonymized perceived race data. In this section, we provide more details about the technical objectives of this system, discussing both privacy and data utility.

Perceived race data is needed to measure the acceptance rate gap. However, the data we use need not be *identifiable*, or attributable to an individual user. Safeguarding our hosts' and guests' data and protecting their privacy is critical to the success of our work. Therefore, we want to only collect and store data at a minimum level of identifiability required to accurately measure discrimination so that, in the event of internal misuse or external leakage of data, users cannot be targeted at scale based upon their perceived race. At this point, we also note that our process of measurement is constrained to only our US communities—the system we've developed to ensure we restrict analysis to only our US communities is outside the scope of this paper.

External leakage of data and abuses of confidential data are well-catalogued phenomena [Shabtai 2012, Sweeney 1997]. Were we merely dealing with confidential demographic data, this alone would warrant heightened rigor and scrutiny. In our context, the bar is raised even higher because:

- We're dealing with particularly sensitive demographic data, race, in a country with a history of racial discrimination and profiling [Rothstein 2017].
- These data are being handled by a large technology company (Airbnb), and the technology industry as a whole deserves increased scrutiny regarding the use (and potential mis-use) of personal data [Hern 2018].

This bar is not meant for all companies nor all contexts—we believe it is most appropriate for companies whose scale of impact merits it and whose resources allow for meeting such a high bar. Our goal in this paper is to empower such companies with the knowledge and tools they need to measure and mitigate experience gaps without having to engage in privacy-by-design from scratch.

The tension between respecting/preserving the privacy of our users and leveraging the benefit of granular data is ever-present in privacy-conscious data analysis and research [AEPD 2019, Sweeney 2000, Mendes 2017]. In the following two subsections, we provide more details on both sides of this tradeoff. We start by discussing more technical details about the privacy threats our system is designed to mitigate, before defining the privacy objective of our system. We then discuss how we measure the usefulness of anonymized data in measuring the acceptance rate gap.

# Disclosure threat categories

For the purposes of this 1 paper, an unauthorized *disclosure* occurs when an *attacker* (also referred to as *misactor*) gains unauthorized access to *sensitive data* [Lambert 1993]. In this subsection, we categorize the different disclosure threats described in the privacy literature. We then conclude this subsection with a precise definition of what we mean when we say the perceived race data in our system is *anonymized*.

We discuss three types of disclosure threats, in order of increasing specificity:

1. *Membership disclosure* occurs when an attacker gains knowledge about whether or not a person's record exists in a dataset [Gkoulalas-Divanis 2014]. Other terms for this include: *table linkage* [Fung 2010].
2. *Attribute disclosure* occurs when an attacker gains knowledge about a sensitive value or values associated with a person.
3. *Identity disclosure* occurs when an attacker gains knowledge about which row or rows in a dataset may correspond to a person [Gkoulalas-Divanis 2014]. Other terms for this include: *singling-out* [Article 29 DPWP 2014], *re-identification* [Lambert 1993], and *unique re-identification* [Sweeney 2018].

See figures 1-3 on the right for examples of each type of disclosure.

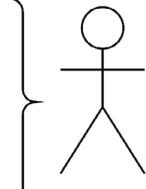

**Figure 1:** Example membership disclosure

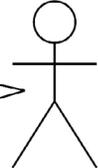

**Figure 2:** Example attribute disclosure

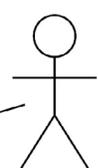

**Figure 3:** Example identity disclosure

Disclosure can vary from *probabilistic* to *certain*: *probabilistic disclosure* occurs when an attacker has increased their knowledge through an attack, but not to the point of being *certain*[2]—this occurs when they are able to revise their posterior beliefs to be largely different from their prior beliefs [Fung 2010, Machanavajjhala 2006]. Additional terms used for various probabilistic disclosures are *inferential disclosure* [HealthStats NSW 2015], *inference* [Article 29 DPWP 2014], *skewness attack* [Domingo-Ferrer 2008, Li 2006], *group re-identification* [Sweeney 2018], and *background knowledge attack* [Machanavajjhala 2006]. See Figure 4 below for an example.

A disclosure may provide an attacker knowledge that may be *correct* or *incorrect*. A disclosure may still cause harm even if incorrect [Article 29 DPWP 2014, Sweeney 2018, Lambert 1993]. In considering disclosure in our system, we do not distinguish between disclosure that is actually correct or incorrect—we aim to mitigate the risk that these data can be used to target individual users based upon perceived race at scale, regardless of correctness.

For our purposes, identifying a user in a dataset (a breach of *anonymity*) is only relevant if doing so leads to knowledge of the user's perceived race (a breach of *anonymity with respect to an attribute* [Sowmyarani 2013]). In other words, we're primarily concerned with attribute disclosure where the sensitive attribute is perceived race. Membership disclosure isn't relevant in our context because it won't lead to certain knowledge of perceived race (this is tied to the p-sensitivity of the dataset, discussed in later sections). Identity disclosure is only relevant insofar as the row or rows contain perceived race, in which case attribute disclosure has also occurred. Our primary focus is thus on preventing certain attribute disclosure—because of this focus (sensitive attribute in lieu of membership, certain in lieu of probabilistic), ε-differential privacy is not an appropriate base privacy model on which to build our system. However, in future work non-interactive forms of differential privacy may be overlaid to mitigate probabilistic disclosure—see the section **Future work: Limiting probabilistic disclosure** [Fung 2010, Mohammed 2011].

The scale of potential disclosure is also relevant. Our paper discusses disclosure with respect to *an individual* for convenience of language, but our goal is really to mitigate the risk of disclosure *at scale* so that a significant number of users cannot be targeted based upon their perceived race. After all, if a misactor is looking to target a relatively small number of users based upon their perceived race, rather than attempting a complicated privacy/security attack, the easiest way to do so would be to manually perceive the race of those users themselves by using available data. In other words, we want to ensure that unauthorized access to the perceived race data doesn't help a misactor to target individuals based upon their race much more than access to the identifiable data (profile photo and first name) alone.

In summary, we consider data to be *anonymized* in this initiative if the risk of *certain sensitive attribute disclosure at scale* (where the sensitive attribute is *perceived race*) is sufficiently mitigated. In the section **System design,** we describe how we enforce the privacy model of *p-sensitive k-anonymity* in order to prevent certain sensitive attribute disclosure; in a later section, **Disclosure risk analysis**, we analyze how the system mitigates the disclosure risks discussed.

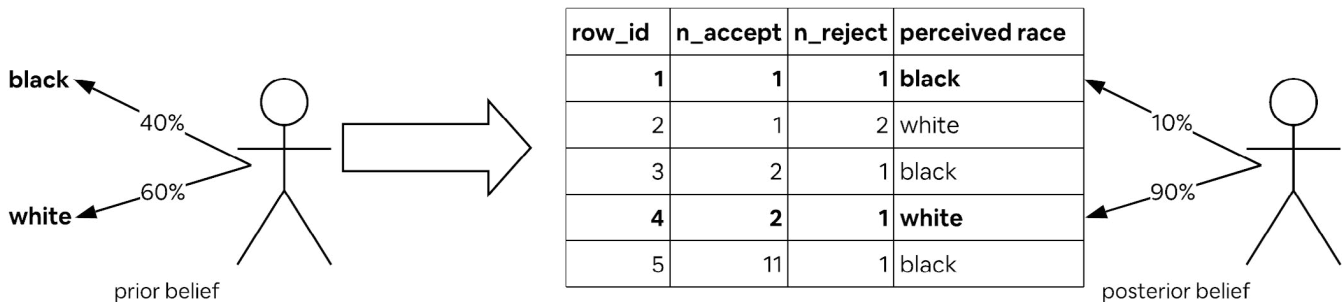

**Figure 4:** Example probabilistic identity disclosure

---

# Data utility

While identifiable data is not necessarily required to measure the acceptance rate gap, the anonymization process could still make accurate measurement difficult or impossible. The literature has shown that anonymization can harm *data utility*, the usefulness of data for analysis [Wu 2013, Daries 2014, Angiuli 2015]. Given the importance of our team's ability to measure progress, we have decided to engage in an in-depth study of how the anonymization process affects data utility. We focus on *task-specific data utility,* where the task we study is measuring the impact of features intended to reduce the acceptance rate gap. We are interested in the effect of anonymization on *statistical power*—the probability that we detect a change if a true change occurred. We develop a simulation-based framework to analyze this relationship. We find that, while enforcing anonymity reduces statistical power through a 0-20% relative increase in minimum detectable effects, we can offset this impact through running studies with larger sample sizes. More details can be found in the section **Simulation construction and results**. The framework we present was designed to be adaptable to other contexts, acting as a supplement to more general measures of data utility that are task-independent.

The anonymized perceived race data described in this paper may help us measure, and consequently work to mitigate, experience gaps beyond the acceptance rate gap. There's no blueprint for combating discrimination on online platforms, so we have to develop an understanding of the problem as it pertains to Airbnb from the ground up. We might use these data, for example, to study patterns in Instant Book usage. Instant Book allows guests who meet a host's requirements book a place to stay immediately without needing the host's individual approval. Since Instant Book leads to a 100 percent initial acceptance rate[3], making it easier for people to use Instant Book can lead to a reduction in the acceptance rate gap. Task-independent measures of data utility, such as those discussed in [Gkoulalas-Divanis 2014], are valuable in ensuring that the anonymized data is useful for such applications.

The rest of the paper is outlined as follows. The next section provides an in depth overview of the **System design**, introducing the privacy model of p-sensitive k-anonymity and the mechanism we use to enforce it. **Simulation construction and results** then discusses the data utility of the anonymized data. The next section, **Disclosure risk analysis,** enumerates the certain sensitive attribute disclosure threats present in the system. We conclude by discussing potential improvements to our system in **Future work**, before summarizing key takeaways in the **Conclusion**.

---

3 Instant Book in its current form is not an effective blanket solution for the acceptance rate gap. Additional internal research has shown that, given concerns about discrimination, Instant Book is not always an attractive or perceived-as-viable method of booking for Black guests.

# System design

In this section, we walk through a simplified data flow diagram of a system that allows our team to, in the final step, measure the acceptance rate gap using anonymized perceived race data. As we walk through each process and its associated datastores, we introduce the relevant privacy models *k-anonymity* and *p-sensitive k-anonymity* and associated anonymization methods.

The main processes in our system are:

**Guest Booking + Select Guest Data:** the process by which US guests book on Airbnb and we collect the relevant data for computing acceptance rates (e.g., whether a reservation request was accepted or rejected).

**K-Anonymize:** this process has two main components:

1. Using a combination of generalization and suppression to achieve k-anonymity for the data used to compute acceptance rates.
2. Preparing data for a *research partner* to perceive race. The *research partner* is an external affiliate who is under a confidentiality agreement with Airbnb and has their systems reviewed by Airbnb security. We use asymmetric encryption to ensure that the internally identifiable data prepared exclusively for and sent to the research partner cannot subsequently be retrieved by Airbnb and re-linked to the k-anonymized data.

**Perceive Race:** our research partner assigns perceived race to pairs of photos and first names. The existence of this separate organization (and associated trust boundaries) allows for the data flow to be one-way to mitigate the risk that it be re-linked to the k-anonymized data by Airbnb.

**P-Sensitize:** we perturb the results of the previous process in order to achieve p-sensitivity prior to storing it.

**Measure ARG:** we use the k-anonymized p-sensitive data to measure the acceptance rate gap (ARG).

Figure 5 below maps out how these processes relate to each other in a Data Flow Diagram (DFD). There are four types of entities in the DFD:

1. *Data stores*, e.g., **User Data**, are databases or files.

2. *Processes*, e.g., **Select Guest Data**, transform input data into output data flows.

3. *Data flows*, e.g., the arrow from **User Data** to **Select User Data**, show how data is moved across the system, either between processes or to/from data stores.

4. *Trust boundaries*, e.g., **Airbnb anti-discrimination Team Trust Boundary**, represent boundaries where the level of trust changes so that, for an actor to have access to data in datastories and access to run processes within a trust boundary, they would need to be authorized appropriately. A misactor may gain unauthorized access to entities within a trust boundary, as discussed further in the more detailed analysis in **Appendix 2: LINDDUN privacy analysis**.

We now examine each process in detail.

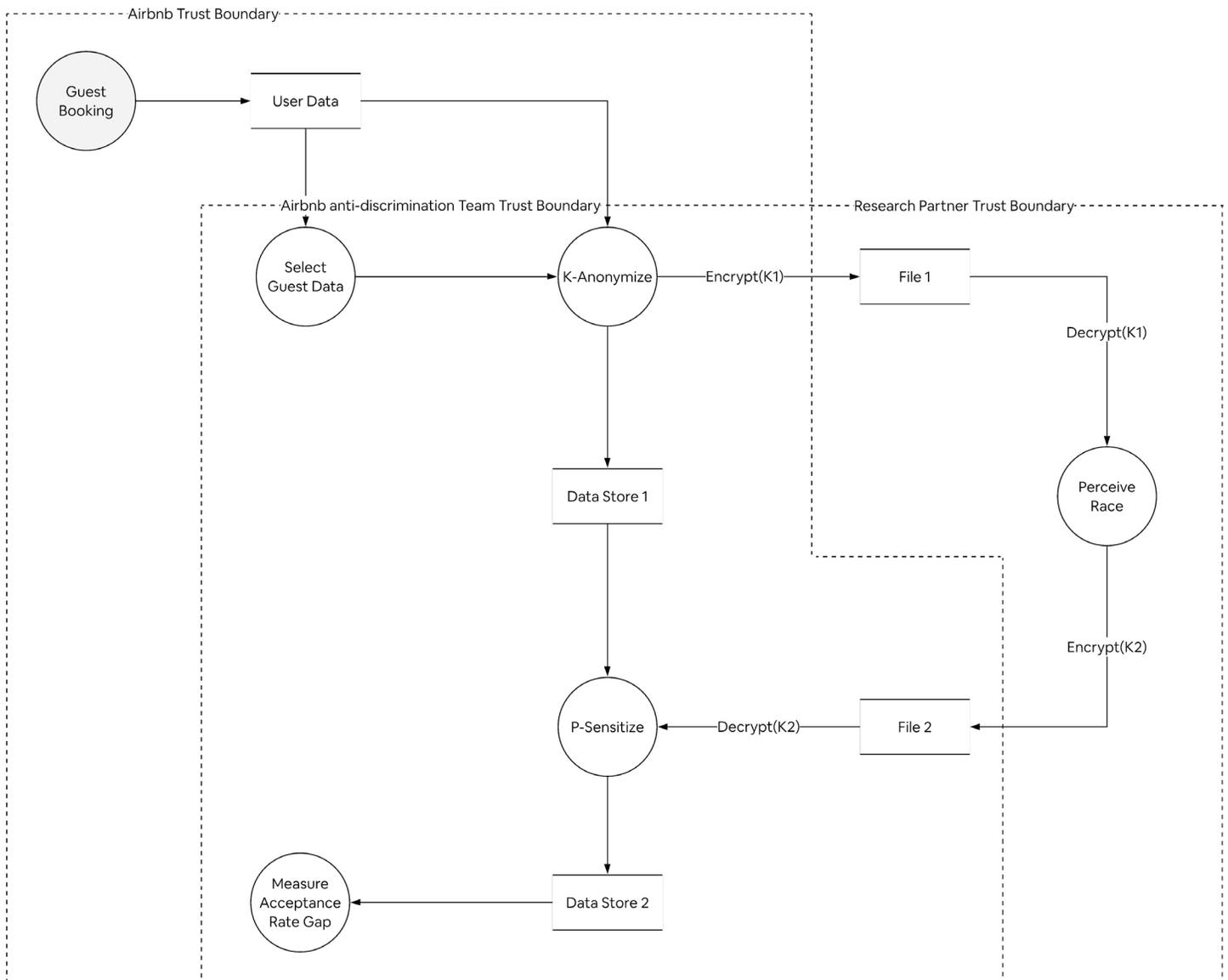

**Figure 5:** Simplified Data Flow Diagram (DFD)

# Guest Booking

**Guest Booking** is the process by which a guest interacts with the Airbnb platform in order to book a stay. This process has been described in the section **Technical Overview**. The resulting data from this process is stored in **User Data** and is available to Airbnb data scientists to conduct analyses. Some examples of the data collected include: which listing a guest tried to book, whether they were accepted, and whether or not they used Instant Book. Details of this process are outside the scope of this paper.

| user_id | n_accept | n_reject |
|---|---|---|
| 1 | 1 | 1 |
| 2 | 1 | 2 |
| 3 | 2 | 1 |
| 4 | 2 | 1 |
| 5 | 11 | 1 |

**Table 1:** Example Select Guest Data output

| user_ids | n_accept | n_reject |
|---|---|---|
| 1 | 1 | 1 |
| 2 | 1 | 2 |
| 3, 4 | 2 | 1 |
| 5 | 11 | 1 |

**Table 2:** Table 1, grouped by equivalence class

# Select Guest Data

The Airbnb anti-discrimination team's data scientist prepares an aggregation of US guests' booking acceptances and rejections to measure the acceptance rate gap. Table 1 on the left provides an example output for **Select Guest Data** that is used in the remainder of this paper. Safeguards and processes to ensure only US guest data are analyzed are outside the scope of this paper.

Membership and identity disclosure threats are inherent in this dataset due to the identifying attribute `user_id`. Furthermore, given knowledge of membership, an attacker may be able to achieve identity disclosure through the other attributes in the dataset. *Quasi-identifiers*[4] are those attributes that may be used to determine the guest associated with a data row through linking to other data [Dalenius 1986]. In the dataset above, `n_accept` and `n_reject` are quasi-identifiers; for instance, if an attacker knew there was only one user (`user_id = 2`) whose aggregation of contacts by **Select Guest Data** would produce `n_accept = 1, n_reject = 2`, then the attacker could uniquely identify that user's row in the dataset, even if the identifying attribute `user_id` were not present.

Recognition of which attributes are quasi-identifiers and, more importantly, recognition of which attributes are **not** quasi-identifiers, cannot be done perfectly due to (1) incomplete knowledge of the data recipients knowledge [Sweeney 1997] and (2) increasing availability of data and capabilities for linkage attacks over time [Cavoukian 2012]. (1) covers to some degree the notion of an internal attacker with access to analytics data on our users [Domingo-Ferrer 2008], as well as an external attacker gaining unauthorized access to our systems via hacking or unauthorized scraping (as considered, e.g., in [Daries 2014]). Additionally, the empirical risks of re-identification may be higher than the standard worst-case scenario considered in k-anonymity, defined in the next section, (1/k) due to additional attributes not considered as quasi-identifiers [Basu 2015].

This has led to experts' recommendation that all non-sensitive attributes be considered at least quasi-identifiers unless explicitly argued otherwise [Sweeney 2018]. Thus, we consider all attributes, besides perceived race, to be quasi-identifiers because an attacker may seek to link back to any dataset, most prominently **User Data**.

---

4 Other terms used for quasi-identifier in the literature include: *key attributes* [Domingo-Ferrer 2008], *indirect key-variables* [Domingo-Ferrer 2002], *q\*-blocks* [Machanavajjhala 2006], and *pseudo-identifiers* [Aggarwal 2005].

# K-Anonymize

The **K-Anonymize** process transforms the dataset from **Select Guest Data** to achieve k-anonymity and prepares data for subsequent processes. In this subsection, we give an overview of k-anonymity[5], walk-through two methods together used to achieve k-anonymity (generalization and suppression), and provide a detailed description of the **K-Anonymize** process.

A group of data rows whose quasi-identifiers have identical values is termed an *equivalence class*[6] [Li 2006]. In Table 2 on the previous page, we group `user_ids` by equivalence class so that there are four equivalence classes, or unique values of the tuple `n_reject, n_accept`.

For a dataset to be *k-anonymous*, besides expunging identifiers (in the example dataset, `user_id`), all equivalence classes must include at least k data rows [Sweeney 2002]. The above dataset, for instance, is k-anonymous with k = 1 because there exist three equivalence classes with 1 row (`user_id`) and one equivalence class with 2 rows. k-anonymity is often achieved through a combination of *generalization* and *suppression* [Samarati 1998]. We walk through these two methods in sequence below.

*Generalization* means replacing attribute values with less precise values. In the example dataset, if we examine the values for `n_accept` and `n_reject` for the equivalence classes for `user_id = 1` and `user_id = 2` in Table 1 on the previous page we see that `n_accept` is already identical; if we replace `n_reject` values of `1` and `2` for these two rows with the closed interval `[1, 2]` then we will have merged the two equivalence classes into one, as shown in Table 3 below.

*Suppression* means suppressing, or removing, rows or attribute values from a dataset; it can be helpful when considering outlier values that, if generalized with other values, would lead to a significant reduction in data utility [Samarati 1998]. For the example dataset, if we suppress the row associated with `user_id = 5` (whose `n_accept` value of `11` is an outlier), we will achieve k-anonymity on the dataset for k = 2, as shown in Table 4 below.

In order to measure the acceptance rate, we will need to transform the quasi-identifier intervals into numbers. In order to maximize data utility, we *microaggregate* the values of certain quasi-identifiers in each equivalence class by taking the arithmetic mean [Domingo-Ferrer 2002]. We discuss potential improvements to better leverage microaggregation in **Future work: Measuring additional experience gaps**. Replacing intervals with arithmetic means yields the 2-anonymous microaggregated dataset shown in Table 5 below.

Equipped with an understanding of k-anonymity, we can now walk through the **K-Anonymize** process. First, **K-Anonymize** transforms the dataset to achieve k-anonymity as described above. Before discarding the Airbnb internal identifier `user_id`, it generates a random row identifier that will be used for subsequent processes; we call this new identifier a `nid`. The mapping between `user_id` and `nid` is no longer available after **K-Anonymize** completes.

| user_ids | n_accept | n_reject |
|---|---|---|
| 1, 2 | 1 | [1, 2] |
| 3, 4 | 2 | 1 |
| 5 | 11 | 1 |

**Table 3:** Table 2, generalization applied

| user_ids | n_accept | n_reject |
|---|---|---|
| 1, 2 | 1 | [1, 2] |
| 3, 4 | 2 | 1 |

**Table 4:** Table 3, suppression applied

| user_id | n_accept | n_reject |
|---|---|---|
| 1 | 1 | 1.5 |
| 2 | 1 | 1.5 |
| 3 | 2 | 1 |
| 4 | 2 | 1 |

**Table 5:** Table 1, 2-anonymized and micro-aggregated

---

5 See **Technical overview: Disclosure threat categories** for a discussion of our choice of k-anonymity, rather than ε-differential privacy, as our base privacy model.

6 Other terms for equivalence class include *P-set* [Narayanan 2008] and *QI-group* [He 2016].

K-Anonymize persists the k-anonymous dataset with `nid` to **Data Store 1**—an example dataset (following the example from above) is shown in Table 6 below. Note that **Data Store 1** does not yet contain perceived race data—that data is introduced in subsequent processes below.

**Data Store 1** is stored securely and data is automatically deleted 30 days after persisting. Access to **Data Store 1** is restricted to authorized members of the Airbnb anti-discrimination team.

At this point, the data scientist may inspect **Data Store 1** before the remaining processes are followed. This allows for additional out-of-band checks on expected data utility prior to initiating the expensive manual process of collecting perceived race data (**Perceive Race**, discussed below).

**K-Anonymize** also creates **File 1** (using `user_id` and **User Data**) for use by the **Research Partner**. **File 1** contains the data required for **Perceive Race**: a identifier for each user row, the user's first name (seen by the host when they accept/reject a guest), and the profile photo for the user. An example dataset (following the example from above) is shown in Table 7 below.

Using a public key provided by **Research Partner**, **File 1** is asymmetrically encrypted prior to persisting so that, because only **Research Partner** has the private key, **Airbnb** can no longer decrypt **File 1** (which is why it lies within the **Research Partner Trust Boundary**) and thus loses the linking from `nid` to `photo_url`. The linkage between `nid` to `user_id` is not persisted so that, because **File 1** cannot be decrypted by Airbnb, it is lost when **K-Anonymize** completes. If Airbnb had access to the data in **File 1**, they could re-link to `user_id` through `nid → photo_url → user_id`. This is discussed in further detail in a later section, **Disclosure risk analysis**.

# Perceive Race

The **Research Partner** is given pairs of first names and photos of Airbnb hosts or guests. First name is provided because studies have shown that first names are a strong signal for perception of race in the United States [Bertrand 2004]. **Appendix 3 Research Partner requirements** outlines a comprehensive list of requirements a **research partner** must satisfy, including regular security review and confidentiality.

They then assign a perceived race to each pair of first names and photos. In our system design, we require the **Research Partner** to engage in the human perception of race, i.e. to not utilize algorithms for computer perception of race. We've imposed this requirement for two reasons:

1. The use of and improvement of algorithms that use facial images for perception of race warrants additional scrutiny and input from, among others, our outside civil rights and privacy partners [Fu 2014].

2. Preliminary internal research has shown that pre-existing algorithms that don't use facial images, such as Bayesian Improved Surname Geocoding (BISG), are not sufficiently accurate to allow for measuring the acceptance rate gap with precision [CFPB 2014].

A later section **Future work: De-identifying user profile photos** mentions a general method currently being internally developed of measuring the impact changes to the inputs for **Perceive Race**, so that we can later more robustly assess the accuracy of pre-existing algorithms such as BISG while still using this overall system and the privacy protections it provides.

| nid | n_accept | n_reject |
|---|---|---|
| 74 | 1 | 1.5 |
| 60 | 1 | 1.5 |
| 73 | 2 | 1 |
| 22 | 2 | 1 |

**Table 6:** Example Data Store 1

| nid | first_name | photo_url |
|---|---|---|
| 74 | Al | https://photo.url/b732d714-108f-4858-9f2e-d08418196464.jpg |
| 60 | Ramon | https://photo.url/e0705d2b-d3f2-4b6e-acd0-9fc667f7c62c.jpg |
| 73 | Luther | https://photo.url/DB0D8816-e432-4dd5-bf58-cfb2059c308f.jpg |
| 22 | Shelia | https://photo.url/42185889-18f4-4316-b7db-2d266e1d4d28.jpg |

**Table 7:** Example File 1

The data defining who to perceive is provided in **File 1**, for which only the **Research Partner** has the private key to decrypt, so that Airbnb doesn't have access to the data in **File 1**. The **Research Partner** creates **File 2** with the results—an example dataset (following the example from above) is shown in Table 8 below.

**File 2** is also asymmetrically encrypted, this time with a public key provided to the **Research Partner** by **Airbnb**; only authorized members of the Airbnb anti-discrimination team have the private key (which is why it lies within the **Airbnb anti-discrimination Team Trust Boundary**).

| nid | perceived_race |
|---|---|
| 74 | white |
| 60 | white |
| 73 | white |
| 22 | black |

**Table 8:** Example File 2

| nid | n_accept | n_reject | perceived_race |
|---|---|---|---|
| 74 | 1 | 1.5 | white |
| 60 | 1 | 1.5 | white |
| 73 | 2 | 1 | white |
| 22 | 2 | 1 | black |

**Table 9:** Example joined Data Store 1 and File 2

| nid | n_accept | n_reject | perceived_race |
|---|---|---|---|
| 74 | 1 | 1.5 | white |
| 60 | 1 | 1.5 | **black** |
| 73 | 2 | 1 | white |
| 22 | 2 | 1 | black |

**Table 10:** Table 9, p-sensitized

# P-Sensitize

**P-Sensitize** joins **File 2** and **Data Store 1**, perturbs `perceived_race` as needed to achieve p-sensitivity (defined below), and persists the dataset (without `nid`) to **Data Store 2**. **File 1** and **File 2** are deleted upon successful persistence to **Data Store 2**. If we join the examples for **File 2** and **Data Store 1** from Table 8 and Table 6, we have the dataset in Table 9 on the left.

While this example dataset achieves k-anonymity for k = 2, sensitive attribute disclosure may still occur in the following scenario (known as a *homogeneity attack* [Basu 2015, Machanavajjhala 2006]):

1. Suppose an attacker knows that `user_id = 1` is in the dataset (membership disclosure, though knowledge of **Select Guest Data**).

2. Through knowledge of the generalization strategy (knowledge of the business logic in **K-Anonymize**) the attacker knows that the user's quasi-identifiers have been generalized into the equivalence class with `n_accept = 1, n_reject = 1.5`.

3. Because all rows in that equivalence class have the same value for the sensitive attribute `perceived_race` (a *homogeneous equivalence class*), the attacker can infer that the `perceived_race` associated with the `user_id = 1` is white.

*p-sensitive k-anonymity*[7] means that each equivalence class (of cardinality >= *k*) in a dataset has at least *p* distinct values for a sensitive attribute [Truta 2006]. In this example, the dataset is 1-sensitive 2-anonymous because the equivalence class `n_accept = 1, n_reject = 1.5` has only 1 distinct value for `perceived_race`: white. If we would like it to be 2-sensitive 2-anonymous, we can perturb (or randomly change) values for `perceived_race` where needed to achieve p = 2, which may result in the following modified dataset in Table 10 on the left.

---

One criticism of p-sensitivity is that it may entail substantially reduced data utility due to the sparsity of distinct values for a sensitive attribute in the original dataset [Domingo-Ferrer 2008]. Another criticism is that, if an attacker understands the generalization strategies used to achieve p-sensitive k-anonymity (in this system, knowledge of the business logic in **K-Anonymize** and **P-Sensitize**) and has background knowledge of the quasi-identifiers (in this system, knowledge of **User Data** and **Select Guest Data**), they may be able to achieve sensitive attribute disclosure for some users through a *minimality attack*[8] [Wong 2007].

**P-Sensitize**, because it utilizes perturbation rather than generalization, circumvents both of these issues: the data utility reduction is limited by the rows perturbed; a minimality attack requires generalization in order to infer homogenous equivalence classes that required further generalization.

Once **P-Sensitize** perturbs the data to achieve p-sensitivity (k-anonymity had already been satisfied by **K-Anonymize**), it strips the `nid` and persists the output to **Data Store 2**—an example dataset (following the example from above) is shown in Table 11 below.

**Data Store 2**, as with **Data Store 1**, is stored securely, data is automatically deleted 30 days after persisting, and access is restricted to authorized members of the Airbnb anti-discrimination team. **P-Sensitize** finally deletes **File 1** and **File 2**.

| n_accept | n_reject | perceived_race |
|---|---|---|
| 1 | 1.5 | white |
| 1 | 1.5 | black |
| 2 | 1 | white |
| 2 | 1 | black |

**Table 11:** Example Data Store 2

# Measure Acceptance Rate Gap

The final step in the system is for the team's data scientist to measure the acceptance rate gap, the experience gap for which we're focusing in this paper, using the p-sensitive k-anonymous aggregate contact data that includes perceived race data in **Data Store 2**. Further details around how we measure the acceptance rate gap and changes in it are provided in the next section, **Simulation construction and results**.

---

8 Special thanks to Dr. Cathy O'Neil and Jacob Appel of O'Neil Risk Consulting & Algorithmic Auditing for constructing a scenario demonstrating a minimality attack.

# Simulation construction and results

In this section, we lay out the methodology we use to understand how anonymizing data affects our ability to measure the acceptance rate gap.

The literature has shown that anonymization can adversely affect data utility [Wu 2013, Daries 2014, Angiuli 2015, Karr 2006, Drechsler 2011, Gkoulalas-Divanis 2014]. Therefore, we have dedicated a substantial amount of time to estimate the consequences of enforcing anonymity on data utility. Our primary uses of anonymized data are calculating the acceptance rate gap and measuring changes in the acceptance rate gap following product launches and other interventions. Airbnb, like most other technology companies, tries to understand the effect of interventions by running A/B tests (where users are randomly assigned to either the A or B group), which are designed to measure changes in outcomes between control and treatment groups [Moss 2014, Overgoor 2014, Parks 2017]. This leads us to study the effect of anonymizing data on task-dependent data utility, where the task we consider is measuring the impact of A/B tests on the acceptance rate gap.

More specifically, we analyze how data anonymization affects the statistical power of these tests, as well as the distribution of the estimates that they yield. We develop a simulation-based framework for this: we vary parameters such as the value of k we use in k-anonymization, how long we run A/B tests for, and the effectiveness of our interventions. We then study the relationship between these factors and our measures of data utility. Simulation-based approaches have previously been used to measure data utility in the statistical disclosure limitation (SDL) literature [Karr 2006, Drechsler 2011]. Our work is novel in applying a simulation-based methodology to understanding the impact of anonymization on the efficacy of A/B tests. Our framework is designed to be easily adapted to suit many contexts where A/B testing is used, such as other large technology platforms where experience gaps may exist undetected.

# Simulation construction

We begin by providing an overview of the simulation we run to measure data utility. Several steps in this process should appear familiar to the reader, as they correspond to steps described earlier in our **System Design**. To make this connection clear, we add the name of the relevant **System Design** section in parentheses next to each step of the simulation. Nevertheless, there are important distinctions between the overall system and the simulation. Firstly, we run the entire data flow many times, varying parameters such as k, in order to study their impact on data utility. Secondly, the simulation has no step to perceive race (**Perceive Race**). Instead, we randomly assign users into groups; the group one is in then affects their probability of acceptance. Finally, we randomly perturb some acceptances and rejections during our analysis in order to mimic the impact of a feature launch that impacted the acceptance rate gap. A more detailed walkthrough of the simulation is below.

## Step 1 (Guest Booking):

We start by simulating data to mimic an experiment on our platform that has had an impact on the acceptance rate gap. We focus on a host-side experiment, which means that each host will be assigned to either the treatment or control group.[9] The data consist of one row per contact sent from guest to host. The dataset will have N (to be determined later) rows and has the following schema:

- *Guest identifier*: a guest can send multiple contacts, so this allows us to discern who sent which contact. This corresponds to `user_id` in the prior section **System design**. As the following process flow shows, it is not necessary for computing an experiment's impact on the acceptance rate gap.
- *Guest group*: we use a pseudo-random number generator to assign guests to group A, B, or C. This corresponds to *perceived race* in the rest of the paper. This is necessary for computing an experiment's impact on the acceptance rate gap. The following analysis focuses specifically on the acceptance rate gap between guests in group A and guests in group B.
- *Host experiment group*: whether the host was in the control or treatment group of the experiment. We use a pseudo-random number generator to determine this, with half of hosts being in control and the other half in treatment. This is necessary for computing an experiment's impact on the acceptance rate gap.

- *Accepted*: indicates whether a contact was accepted or not. We use a pseudo-random number generator to determine this. This is necessary for computing an experiment's impact on the acceptance rate gap. There are two factors that affect one's acceptance probability:
  - The guest's group: we model guests in group A to have the highest probability of acceptance, followed by guests in group B and C.
  - The host's treatment: if a host is in treatment, their acceptance rate of guests in group A does not change. However, their acceptance rates of guests in groups B and C increase by an *effect size* that we control as an input to the simulation. Guests in group B see their acceptance probabilities increase by the *effect size* we see in the following analyses, while guests in group C see an increase that's half the magnitude of *effect size*.

## Step 2 (Select Guest Data):

Once we have generated the contact-level (one row per contact) dataframe, we collapse it into a guest-level (one row per guest) dataframe that's ready for our anonymization processes (Step 3 and Step 4 below). This dataframe has the schema:

- *Guest identifier*
- *Guest group*
- *n_accepted_contacts_treatment:* the number of contacts sent to hosts in the treatment arm of the experiment that were accepted
- *n_rejected_contacts_treatment:* the total number of contacts sent to hosts in the treatment arm of the experiment that were rejected
- *n_accepted_contacts_control:* the number of contacts sent to hosts in the control arm of the experiment that were accepted
- *n_rejected_contacts_control:* the total number of contacts sent to hosts in the control arm of the experiment that were rejected

---

9 This is distinct from a guest-side experiment, where randomization occurs at the guest level and each guest is assigned to either the treatment or control group.

### Step 3 (K-Anonymize):

We then use ARX API, an open source tool for anonymizing datasets [Prasser 2014], to k-anonymize the dataframe. Optimal (achieving the best data utility) k-anonymity is NP-hard [Meyerson 2004]; ARX utilizes a heuristic algorithm to achieve k-anonymity through a combination of generalization and suppression.

### Step 4 (P-Sensitize):

We then p-sensitize the dataframe: for each equivalence class that violates p-sensitivity with p = 2, we randomly select a row and perturb the guest group. Prior to perturbing the data to achieve p-sensitivity, we record the percentage of rows that are part of a homogeneous equivalence class. This is used to assess the potential impact of homogeneity attacks in the **Disclosure risk analysis** section below.

### Step 5 (Measure acceptance rate gap—generate contact data):

We then expand the p-sensitive k-anonymous dataset to be a contact-level dataset. We do this by adding a row for each contact in the p-sensitive k-anonymous dataset, which contains the host's experiment group, the guest's group, and whether the contact was accepted or rejected. For contact counts that are non-integral we use the non-integral component as a pseudo-random weighted coin toss for whether to include an additional contact accept or reject.

### Step 6 (Measure acceptance rate gap—compute experiment result):

We estimate the impact of the experiment by running a regression of the form:

```
accepted ~ a * experiment_group + b * guest_group + c * guest_group * experiment_group
```

Here is a detailed breakdown of the variables shown above:

- `accepted` is 1 if a contact is accepted and 0 otherwise
- `experiment_group` is either control or treatment, depending on which arm of the experiment a host was in
- `guest_group` refers to the demographic group a guest was in. We limit our analyses to guests in groups A and B.

We record `c`, the coefficient on `guest_group * experiment_group`, and whether it's statistically significant or not. In the remainder of this paper, we refer to `c` as the *experiment impact on the acceptance rate gap*.

In this step we also measure privacy metrics used in sections below.

### Run Simulation:

We repeat **Steps 1-6** 1,000 times for each combination of the following experiment setups:

- *k* = 1, 5, 10, 50, 100
- Number of contacts in analysis, *N* = 150,000, 200,000, 250,000, 300,000, 350,000, 400,000, 450,000, 500,000, 550,000, 600,000
- The expected experiment impact on the acceptance rate gap, *effect size* = 1.00, 1.25 1.50, 1.75, 2.00, 2.25 percentage points

For each experiment setup, we can then compute the fraction of the 1,000 simulation results where the null hypothesis of no experiment impact on the acceptance rate gap was rejected.[10] The fraction of the tests that reject the null hypothesis is the *power* of the test. We also compute the smallest effect size for which we have a power of 80 percent as the *minimum detectable effect* of the test.

This framework also allows us to study the impact of k and N on statistical power and the distribution of effects we observe. The results of this analysis are in the next section.

---

10 This occurs when the `c`, coefficient on `experiment_group * guest_color`, is different from zero in a statistically significant way.

# Overall simulation results

Once we have the results from the simulation, the first thing we want to do is analyze the impact of changing k and N on statistical power. Statistical power is our primary measure of data utility, as it measures how effective we can be at detecting the impact of feature launches on the acceptance rate gap. We plot the relationship between effect size and statistical power for various values of N in Figure 6 below.

As expected, statistical power increases with sample size. We also see that power decreases as k increases. This decrease is on the order of magnitude of 1-2 percent when k = 5, but increases to 5-10 percent when k = 100.

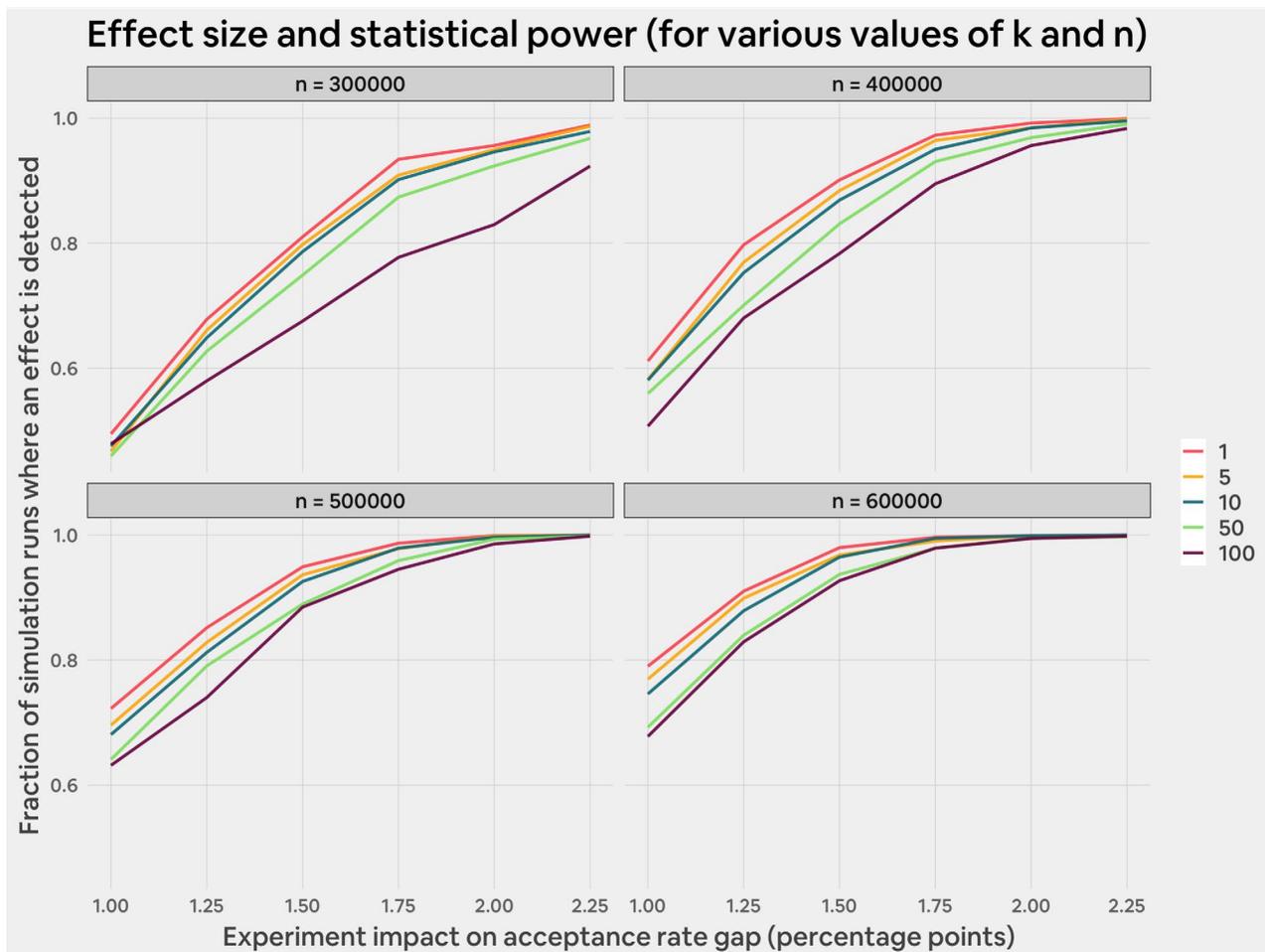

**Figure 6:** Simulation results regarding statistical power

Figure 7 below gives an aggregate view of the above analysis by plotting the minimum detectable effect (smallest effect size with at least 80% power) by N and k. For example, setting k to be 5 or 10 leads to a 0.00 to 0.25 percentage point increase (0-20% relative increase) in minimum detectable effect. Setting k to 100 leads to a 0.25 to 0.50 percentage point increase in minimum detectable effect.

The main takeaway is that anonymizing data increases an experiment's minimum detectable effect by 0-20% (depending on the value of k) in our context. This implies that we can detect the impact of experiments on the acceptance rate gap with anonymized demographic data. However, there is clearly some degree of reduced data utility, or *information loss*, that occurs in the process of enforcing anonymity. Our simulations demonstrate that we can mitigate this loss' effect on statistical power by obtaining a larger sample size (number of contacts). Practically, this means that we can mitigate the effects of data anonymization by running experiments for longer.

There could be other consequences of this information loss on measurement. To dig further into this subject, we analyze the impact of our simulation parameters on the distribution of observed effect sizes.

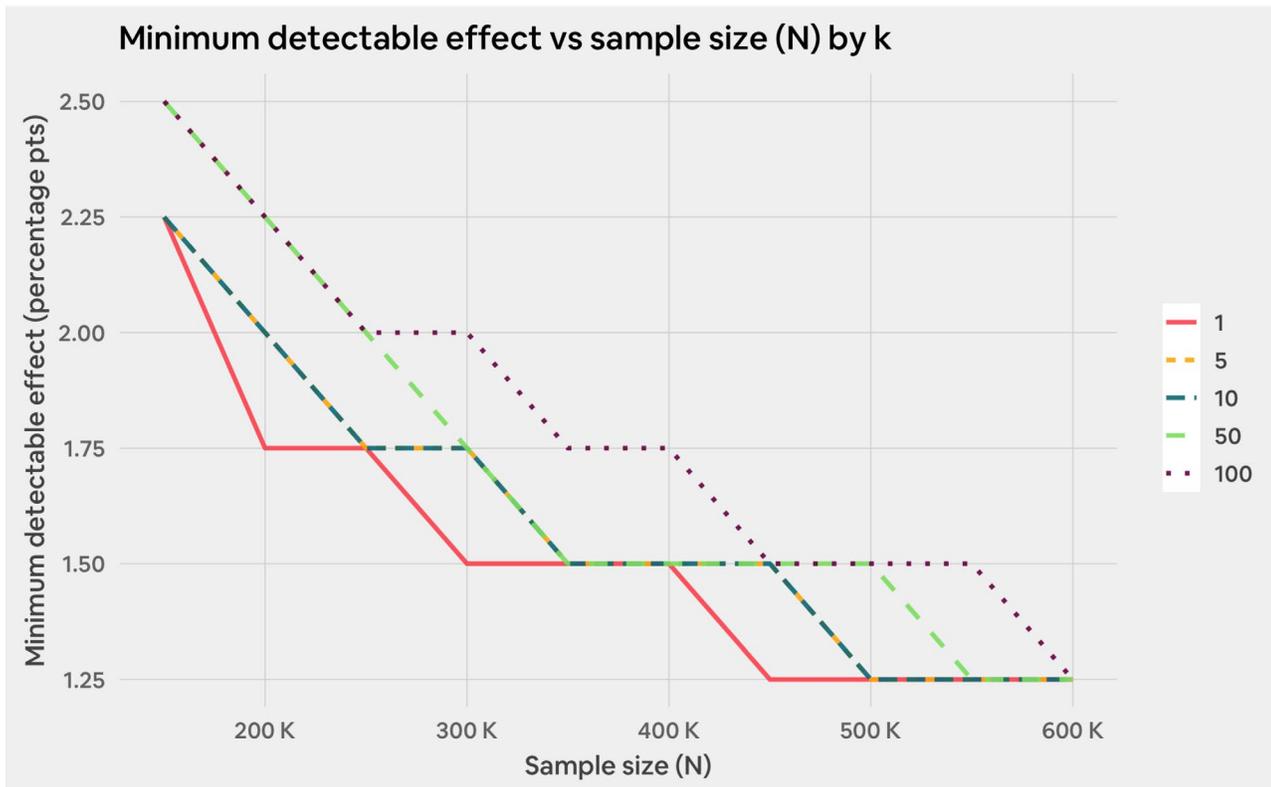

**Figure 7:** Minimum detectable effect

# Initial analysis of the distribution of observed effect sizes

In this section, we study the relationship between k and the distribution of *observed effect sizes*, or the magnitude of the change in the acceptance rate gap we observe in a simulation run. This is equal to the coefficient `c`, on `guest_group * experiment_group` in the regression we run in step 6 of the simulation construction, described above. This is distinct from the expected impact on the acceptance rate gap, which we control as an input to the simulation.

We can gain a more detailed view of how k affects measurement by fixing an experiment size (N) and expected effect size, while also looking at the distribution of observed effect sizes for various values of k. Figure 8 below shows such a distribution for our simulation runs with 400,000 contacts and an expected reduction in the acceptance rate gap of 1 percentage point.

If we did not enforce anonymity, we would expect the distribution of observed effect sizes (c) to be normally distributed with a mean of the expected effect size. This is what we see when k = 1 in Figure 8. When we increase k, the data still appear to be normally distributed. A QQ plot of these distributions (Figure 9 on next page) does not show much evidence for non-normality. However, the distributions appear to be shifted rightwards (i.e. have downward bias) and have higher variance for larger values of k. This implies that our estimates of experiment impact on the acceptance rate gap become less precise as we increase k. The combination of downward bias and higher variance appear to be the principal drivers of reduced statistical power when k is large. We explore this further in the next subsection. Table 12 below provides summary statistics of the simulation runs we conducted that provide more concrete evidence supporting these observations.

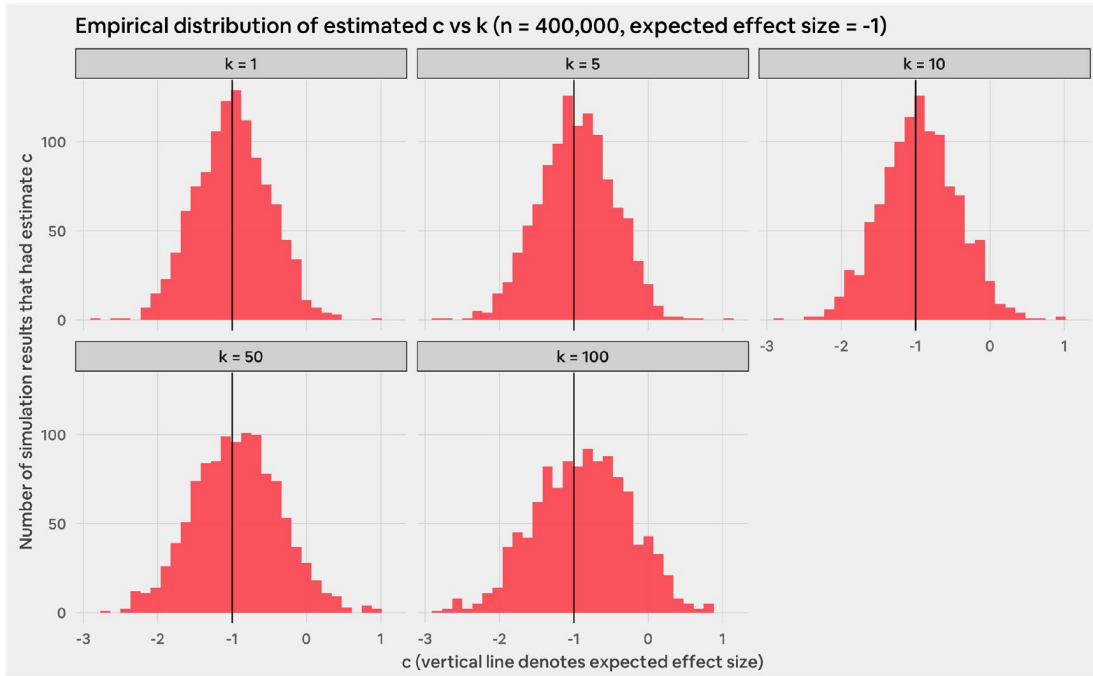

**Figure 8:** Distribution of observed effect sizes by k

| effect_size | k | n_sim_runs | mean_c | bias_pct | sd_c | skew_c | ks_stat | ks_pvalue |
|---|---|---|---|---|---|---|---|---|
| -1 | 1 | 1112 | -0.9913400 | -0.8660022 | 0.4907050 | 0.0182670 | 0.0138169 | 0.9837161 |
| -1 | 5 | 1112 | -0.9663113 | -3.3688709 | 0.5044472 | -0.0345632 | 0.0134261 | 0.9881110 |
| -1 | 10 | 1112 | -0.9450707 | -5.4929328 | 0.5171831 | 0.0546914 | 0.0145551 | 0.9725423 |
| -1 | 50 | 1112 | -0.9343867 | -6.5613342 | 0.5801732 | 0.0708445 | 0.0150543 | 0.9626442 |
| -1 | 100 | 1112 | -0.8866447 | -11.3355278 | 0.6350208 | -0.1008769 | 0.0183106 | 0.8499479 |

**Table 12:** Summary statistics of simulation runs (N = 400,000, expected effect size = -1)

The columns mean_b, sd_b, and skew_b in Table 12 record the mean, standard deviation, and skewness of the observed effect sizes. As suggested by the plots in Figure 8, the mean of the observed effect sizes is smaller than the actual effect size we simulated. This leads to an attenuation bias that ranges from 3.4 percent (k = 5) to 11.3 percent (k = 100). The standard deviation of observed effect sizes also increases from 0.49 (k = 1) to 0.64 (k = 100). The distribution of observed estimates is relatively symmetric and does not show much skew.

We also run a Kolmogorov-Smirnov test of the observed distribution versus a normal distribution with mean and standard deviation equal to those seen in the simulation runs. The columns ks_stat and ks_pvalue in Table 12 record the test statistic and p-value respectively. We do not see much evidence that would lead us to reject the null hypothesis of normality.

Table 13 below shows the same summary statistics as Table 12, for N = 400,000 and an expected effect size of 2 percentage points. The trends we see are similar, with the difference that the magnitude of downward bias is slightly smaller.

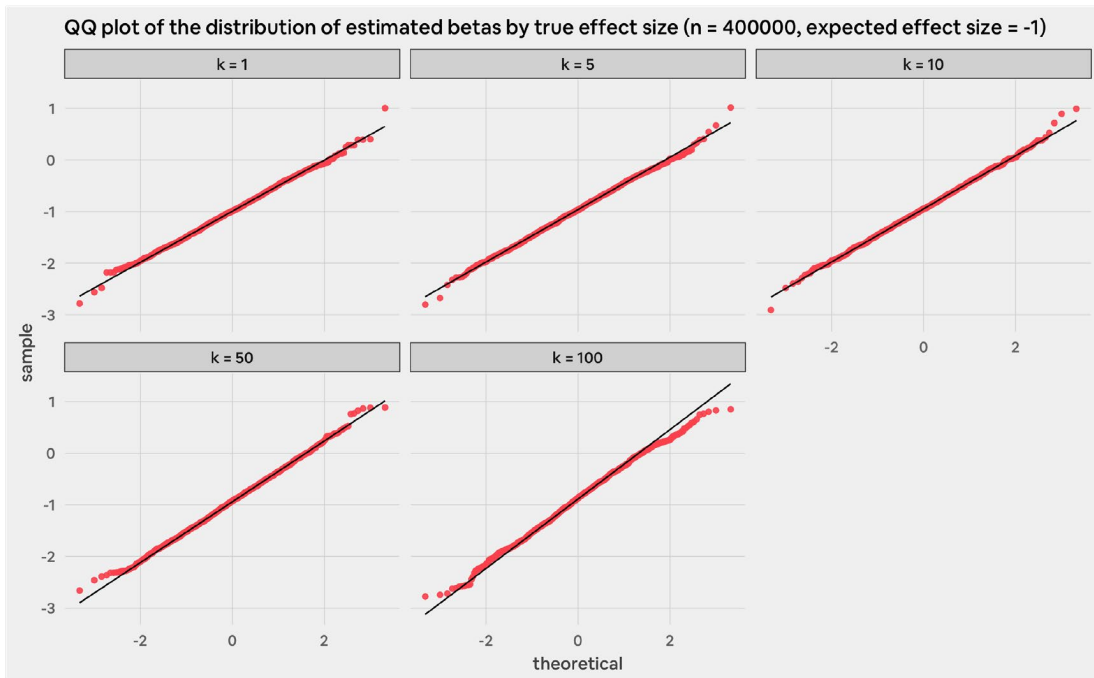

**Figure 9:** QQ plot of observed effect sizes

| effect_size | k | n_sim_runs | mean_c | bias_pct | sd_c | skew_c | ks_stat | ks_pvalue |
|---|---|---|---|---|---|---|---|---|
| -2 | 1 | 1027 | -2.016334 | -0.8166805 | 0.4839818 | 0.0858513 | 0.0209092 | 0.7602996 |
| -2 | 5 | 1027 | -1.960318 | -1.9841184 | 0.4968109 | 0.1061983 | 0.0253950 | 0.5218151 |
| -2 | 10 | 1027 | -1.940240 | -2.9879880 | 0.5164208 | 0.0347253 | 0.0234036 | 0.6271495 |
| -2 | 50 | 1027 | -1.909743 | -4.5128548 | 0.5626589 | 0.0270892 | 0.0205542 | 0.7784200 |
| -2 | 100 | 1027 | -1.839587 | -8.0206474 | 0.5936817 | -0.1023239 | 0.0183405 | 0.8800677 |

**Table 13:** Summary statistics of simulation runs (N = 400,000, expected effect size = -2)

# Further analysis of the bias and variance of our simulation results

We can use the breadth of the simulations we ran in order to improve our understanding of the drivers of statistical bias and variance of our simulation results. We ran over 1,000 simulations for each possible combination of k, N, and expected effect size. This allows us to take a given k and use the results for various N's and expected effect sizes in order to obtain a distribution for statistical bias for that k. Repeating this exercise for various k's allows us to gain a heuristic for the relationship between k and bias.

The first panel of Figure 10 below shows how bias clearly increases as k increases. We can repeat the same exercise for the number of contacts in an experiment and the expected reduction in the acceptance rate gap (second panel of Figure 10). We do not see a large correlation between N and bias, but we do see that bias seems to decrease slightly for larger effect sizes (third panel of Figure 10).

Figure 11 below replicates this analysis for the empirical variance of estimated effect sizes. We see that dispersion increases as k increases and decreases as N increases. Neither of these observations should be particularly surprising. We do not see a major correlation between dispersion and effect size.

Based on this analysis, we believe we can use p-sensitive k-anonymous perceived race data, for p = 2 and k = 5, to measure the impact of different Airbnb features on the acceptance rate gap. We provide additional context and guidelines for selecting k in **Appendix 1: survey of values of k**.

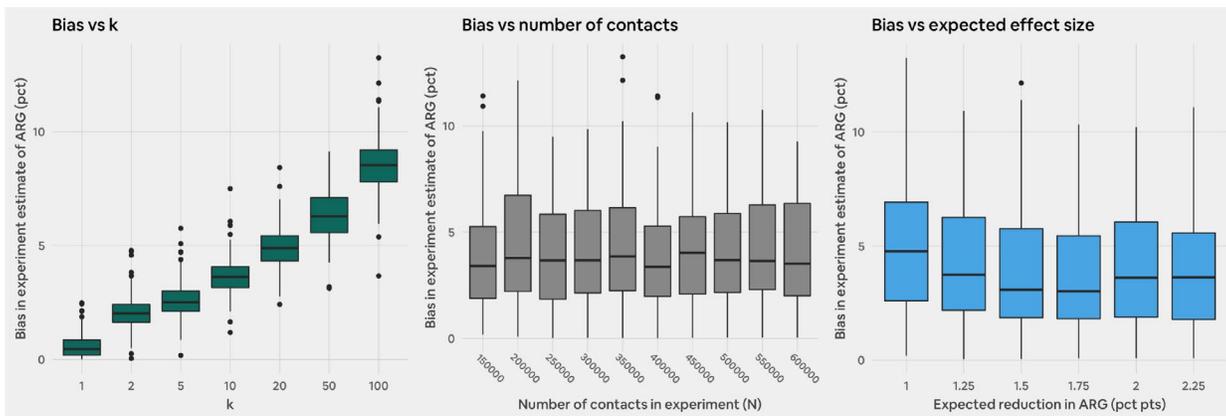

**Figure 10:** Determinants of statistical bias

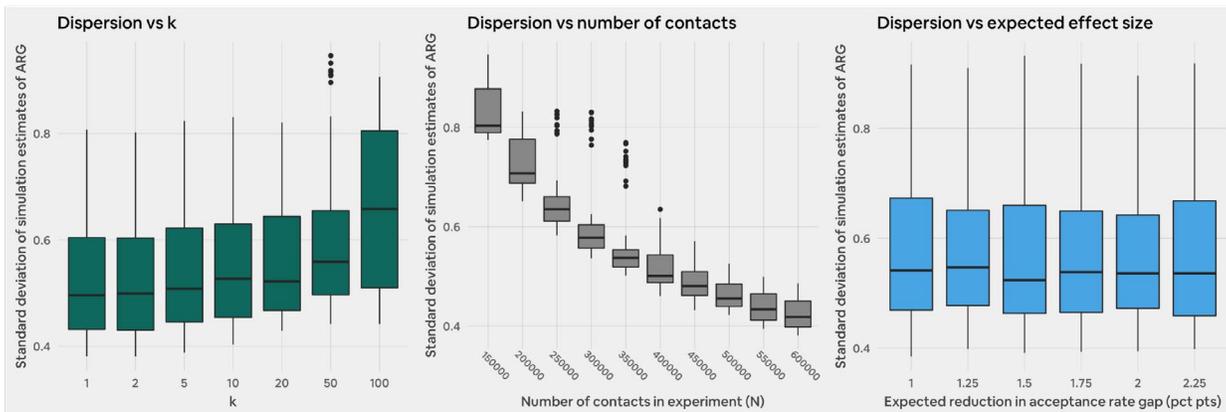

**Figure 11:** Determinants of empirical variance of estimated effect sizes

# Summary of findings from simulations

The major findings from our simulations are summarized below. Firstly, when we enforce k-anonymity with k = 5, we see a 0-20 percent increase in the minimum detectable effect of an A/B test's impact on the acceptance rate gap. We can mitigate this reduction in statistical power by running A/B tests for longer in order to obtain larger sample sizes. After analyzing the distribution of observed effect sizes, we see that this increase in minimum detectable effects is driven by a downward bias and increased variance in our estimates of A/B test impact for higher values of k. We expect that further enhancements to the parameters and algorithms used in the anonymization processes (**K-Anonymize** in particular) will reduce the impact of anonymization on our ability to measure the acceptance rate gap. This may allow us to use higher values of k as well as potentially use this system to measure other potential experience gaps where anonymization may have a larger impact on data utility. We discuss some of these potential enhancements later in the **Future work** section below.

# Disclosure risk analysis

In this section, we enumerate the certain attribute disclosure threats in the system by examining data linkages. We then, for each threat, provide a risk assessment to establish that we believe we have sufficiently mitigated the risk of certain attribute disclosure so that we may consider the perceived race data as anonymized.

**Appendix 2** provides a formal privacy analysis using the LINDDUN privacy threat modeling methodology [Deng 2011]. Though the LINDDUN methodology is helpful in providing a high-level framework for threat modelling, it does not provide tools to help ensure all concrete threats to linkability of data (one of the many categories of threats the methodology attempts to capture) have been explicitly reviewed. In this section, we create and analyze a *data-linkage diagram*, a variation on an entity-relationship diagram, to enumerate all such threats.

In the data-linkage diagram in Figure 12 below, a block represents a datastore[11] from the system design: the title of the datastore is in bold, followed by the relevant data attributes in the datastore. There is an additional block labeled **Public Data,** which represents potentially extant data about users outside the system. Datastores that contain externally identifiable user data are colored yellow[12] (e.g., **User Data** on its own contains sufficient data about most users to uniquely identify them); those that contain perceived race, the relevant sensitive attribute for disclosure analysis, are colored red[13] (e.g., **File 2** contains perceived race). The diamonds describe data linkages (e.g., **Data Store 1** may be related, or linked to, **File 2** via the NID). Solid lines represent data or data linkages that are certain to exist. Dashed lines represent linkages where data may exist (e.g., not all users may have photos available in **Public Data**) or linkages that are probabilistic (e.g., k-anonymous so that the probability of linkage is at-most `1/k`, rather than `1.0`).

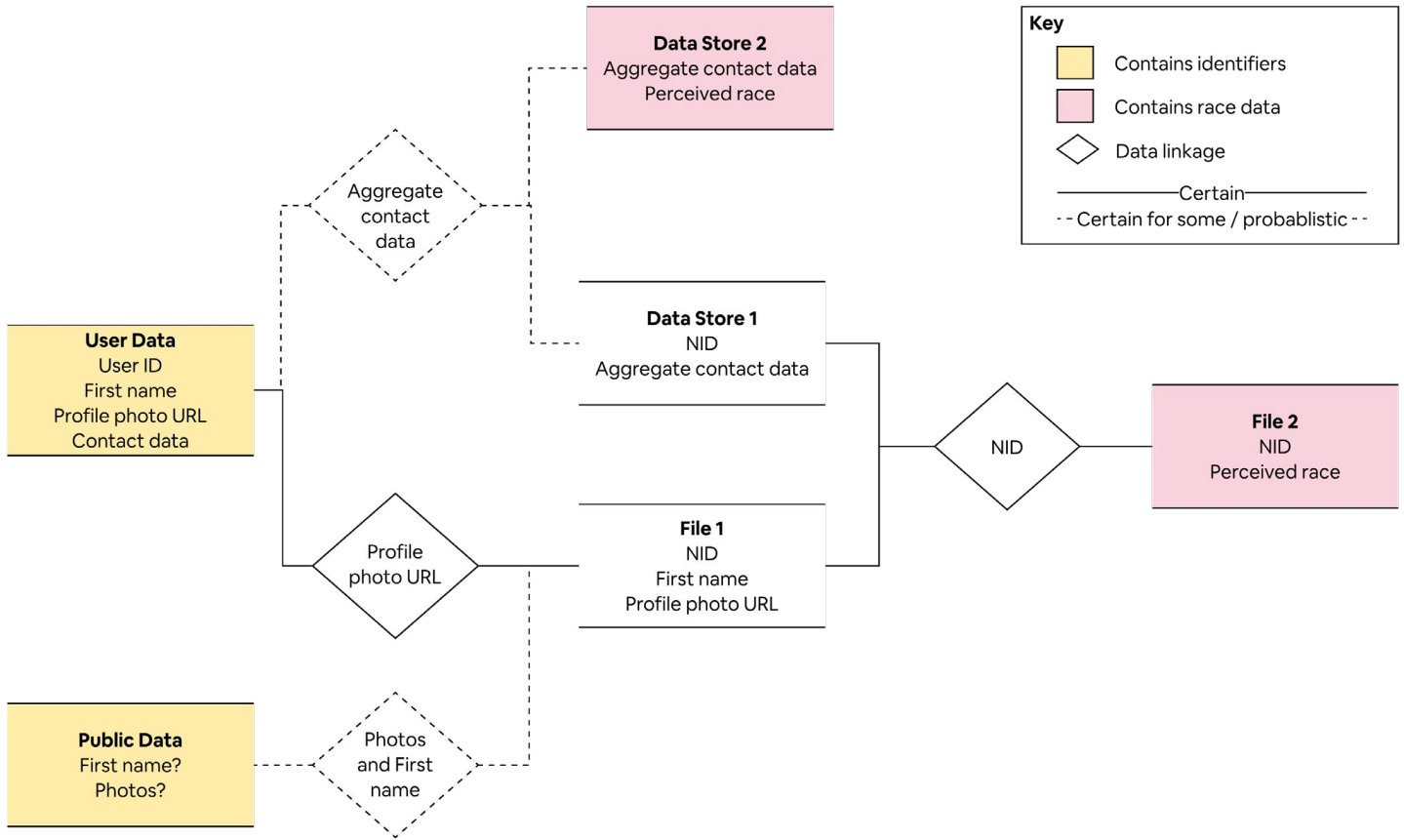

**Figure 12**: Data-linkage diagram

---

11 In this section vs. Appendix 2, we do not distinguish between a datastore and its input/output dataflow—for simplicity all are represented datastores.

12 The following blocks are colored yellow: **User Data**, **Public Data**.

13 The following blocks are colored red: **Data Store 2**, **File 2**.

Each data linkage is described below:

- **User Data ~ Data Store 1, Data Store 2**: given knowledge of the **Select Guest Data** process, aggregate contact data in **User Data** may be matched to that in either **Data Store 1** or **Data Store 2**. However, because aggregate contact data attributes are quasi-identifiers and both data stores are k-anonymous, a user's aggregate contact data in **User Data** may be mapped to no less than k rows in either **Data Store 1** or **Data Store 2**; the probability of each row being matched correctly (certain attribute disclosure) is `1/k`.

- **User Data ~ File 1**: a user's profile photo URL in **User Data** is unique, so that data in **File 1** may be linked to **User Data** using that URL.

- **Data Store 1, File 1 ~ File 2**: the unique row identifier NID (generated by **K-Anonymize**) may be used to link rows between **Data Store 1**, **File 1**, and **File 2**.

- **Public Data ~ File 1**: facial recognition/photo matching may be used to link the user's publicly available photos and the user's profile photo. Unless the exact same photo exists in both **Public Data** and **File 1**, the matching will be uncertain, i.e. probabilistic. First Name from **File 1** may be used to increase certainty through either (1) its direct existence in **Public Data**, or (2) linkage between other quasi-identifiers in **Public Data**, e.g., location data, with publicly available name distributions such as those in public voter registrations.

Linkages from **Public Data** to **User Data** aren't considered because, were an attacker to achieve linkage between **User Data** and a datastore containing perceived race data, linkage to **Public Data** would be unnecessary to achieve the objective of attribute disclosure. It may, however, improve posterior beliefs, i.e. increase the risk of probabilistic attribute disclosure.

Equipped with this data-linkage diagram and an understanding of each linkage, we then enumerate all paths from externally identifiable user data (yellow blocks) to perceived race data (red blocks). Each such path represents a linkability threat in our privacy analysis in Appendix 2. In this section, we review each threat that may lead to certain attribute disclosure[14]. The scenarios are ordered by our subjective prioritization of their risks—the likelihood of the attack scenario and its impact [Deng 2011]. This prioritization will help guide future improvements to system design such as those discussed in the later section, **Future work**.

---

14 Appendix 2 reviews both probabilistic and certain attribute disclosure threats, though doesn't exhaustively examine malicious modification of processes (which this section does examine in some detail). This section only covers certain attribute disclosure. As before, we do not distinguish between a datastore and its input/output dataflow—for simplicity all are represented as a datastore.

# Homogeneity attack

In this scenario, represented in Figure 13 below, an attacker effectively circumvents **P-Sensitize** to engage in a homogeneity attack [Basu 2015, Machanavajjhala 2006]; the attacker achieves certain attribute disclosure for those users who are members of homogenous equivalence classes. Once the attacker joins **File 2** and **Data Store 1** (using NID as a key), they examine all equivalence classes and, for those that have only a single distinct perceived race value, have achieved certain attribute disclosure.

We believe the likelihood of this threat is low as it requires the misactor be a malicious authorized member of the Airbnb anti-discrimination Team or to circumvent access controls to gain access to the datastores. **Future work: Modifying trust boundaries to further protect anonymity** describes potential improvements to further reduce the likelihood of this threat. We believe the impact of this threat is medium—reflecting the small percentage of users we expect, based on our simulations, to be in homogenous equivalence classes (less than 1% for `k >= 5` across most simulations)—see Figure 14 below.

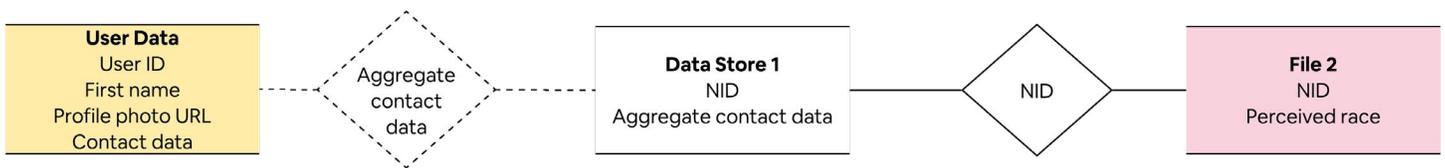

**Figure 13**: Homogeneity attack data-linkage diagram

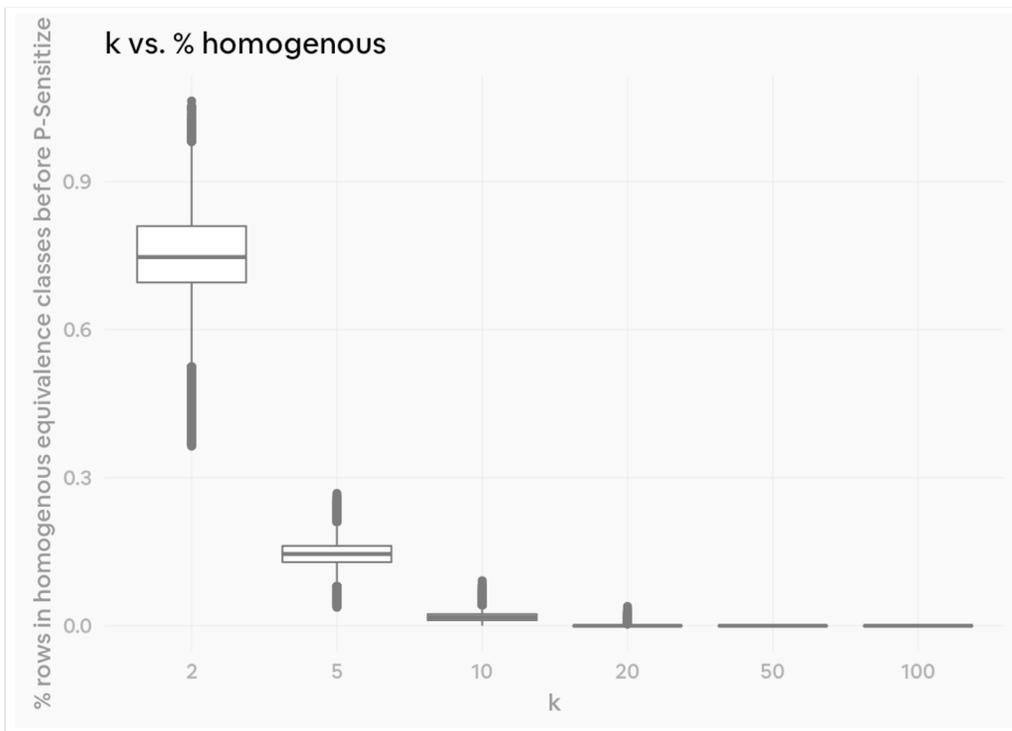

**Figure 14**: Percentage of simulated rows in a homogenous equivalence class for each value of k

# NID ~ User ID mapping attacks

In this scenario, represented in Figure 15 below, an attacker achieves the mapping from NID to User ID using one of three methods (in decreasing order of likelihood):

1. An internal attacker (most likely a malicious authorized member of the Airbnb anti-discrimination team) modifies **K-Anonymize** to capture the NID ~ User ID mapping.

2. An internal attacker circumvents access controls to **Research Partner** to obtain the decryption key for **File 1**, and then joins **User Data** and **File 1** (using the profile photo URL unique to each user) to achieve NID ~ User ID mapping.

3. An external attacker (most likely a malicious authorized **Research Partner** employee) achieves unauthorized access to **User Data** (breaching **Airbnb Trust Boundary**) and knowledge of **Select Guest Data**.

The attacker would then join the resulting User ~ NID data to **File 2** (using NID) to achieve certain attribute disclosure for all users in the dataset.

We believe the likelihood of this threat is low as it requires either a malicious authorized member of the Airbnb anti-discrimination team and/or circumventing access controls across one or more trust boundaries. **Future work: Modify trust boundaries to further protect anonymity** describes potential improvements to further reduce the likelihood of this threat. We believe the impact of this threat is high as it leads to certain attribute disclosure for all users in the dataset.

# Photo matching attack

In this scenario, represented in Figure 16 below, an attacker, likely a malicious authorized **Research Partner** employee, joins **File 1** to **File 2** (using NID) to achieve attribute disclosure for each user whose profile photo from **File 1** exists in **Public Data**.

We believe the likelihood of this threat is high—while we assume the likelihood of internal malicious attackers to be low, externally we assume high. **Future work: De-identifying user profile photos** describes improvements currently being developed to further reduce the likelihood of this threat. We believe the impact of this threat is medium as it leads to certain or probabilistic (depending upon the photo matching method) attribute disclosure for some users in the dataset [Acquisti 2014].

The **Future work** section below maps out areas of future research, some of which may help further mitigate the disclosure risks described above.

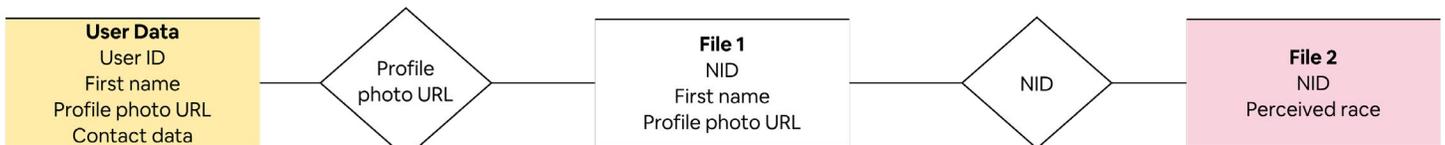

**Figure 15**: NID - User ID mapping attacks data-linkage diagram

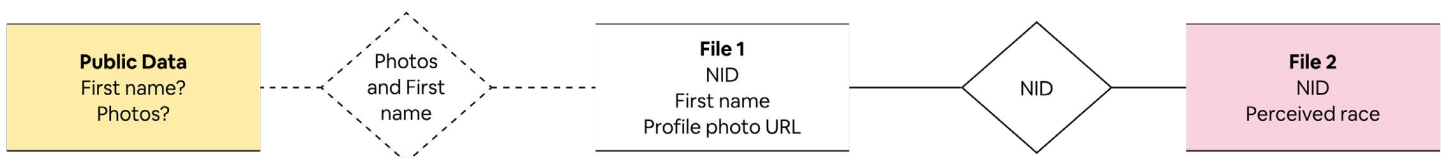

**Figure 16**: Photo matching attack data-linkage diagram

# Future work

In this section, we describe ways to improve and extend our system for measuring discrimination; following this section we conclude with an overview of our findings. These areas of work fit into three broad categories: improving measurement (through increased data utility), improving privacy (through further mitigating disclosure risks), and expanding scope (through measuring additional experience gaps).

# Modifying trust boundaries to further protect anonymity

One area of future work is to investigate modifying trust boundaries to further reduce the likelihood of threat scenarios related to **Homogeneity attack** and **NID ~ User ID mapping attacks** described in the prior section **Disclosure risk analysis**. Two potential modifications of trust boundaries (see prior section **System design**) warrant further analysis:

- Moving **P-Sensitize** to within **Research Partner Trust Boundary**.
- Splitting **Airbnb anti-discrimination Team Trust Boundary** into two separate trust boundaries (with associated organizational firewalls) to ensure that no individual is a member of a team with access to both **K-Anonymize** and **P-Sensitize**.

Each potential change will require careful cross-functional analysis to understand its impact on privacy, data utility, and organizational overhead.

# De-identifying user profile photos

Modifying guests' profile photos to remove identifying (mostly facial) features for use by the **Research Partner** may reduce the likelihood of threat scenarios related to photo matching attacks described in the disclosure risk analysis.

There is a growing body of research on *de-identifying* face images [Alexander 2003, Newton 2005, Gross 2006(a), Gross 2006(b), Muraki 2013]. We will, however, need to measure the impact of various de-identification algorithms on our ability to measure experience gaps because they will impact the accuracy of perception of race by individuals (the **Research Partner**) [Li 2017, Gross 2006(a)].

Because of our risk assessment (see **Disclosure risk analysis: Photo matching attack**), we have begun this research internally and, besides pursuing de-identification of images, are also developing a general method of measuring the impact changes to the inputs for, or to the process of, **Perceive Race** (e.g., de-identification of face images, removal of first name[15], and multiple perceivers) have on accuracy of perception, all while still using p-sensitive k-anonymous data.

# Limiting probabilistic disclosure

One potential area of future work is to enforce additional privacy models to limit probabilistic attribute disclosure. One way to quantify the risk of probabilistic disclosure is *t-closeness*, a metric that measures the amount of information an attacker can gain through knowledge of a user belonging to a particular equivalence class beyond the knowledge gained through overall membership disclosure alone [Li 2006]. In the future, we may pursue enforcing t-closeness or its variants [Li 2006, Soria-Comas 2015, Sowmyarani 2013]; other more nuanced definitions of *l-diversity* beyond distinct l-diversity (a synonym for p-sensitivity) [Machanavajjhala 2006]; other methods such as *(α, k)-anonymity* [Wong 2006]; and the introduction of uncertainty to the generalization procedure[16], perhaps as part of non-interactive forms of differential privacy [Mohammed 2011].

---

15 Special thanks to Gennie Gebhart of the Electronic Frontier Foundation for pointing out the increased photo matching attack risk for users whose first name is unique—this led us to expand this research project to prepare for future increased mitigation of this risk for those users.

16 Special thanks to Dr. Cathy O'Neil and Jacob Appel of O'Neil Risk Consulting & Algorithmic Auditing for suggesting the introduction of noise to the first anonymization process (**K-Anonymize**).

# Improving data utility

There are several approaches that we can consider to improve data utility in measuring experience gaps. Firstly, we can draw from the literature and pursue an alternative to generalization as a method to achieve k-anonymity. One such alternative is *anatomization*, where individual row quasi-identifiers are not generalized and are instead decoupled from sensitive attributes [Xiao 2006]. Another alternative is *semi-homogenous generalization,* an extension of generalization whereby a data row may be a member of multiple equivalence classes [He 2016].

Rather than pursuing an alternative to generalization, we may also consider prioritizing the order in which quasi-identifiers are generalized to better maximize data utility. An example where this method could be helpful is to use pre-experiment data to reduce variance in our estimates of the acceptance rate gap. In our context, we can use features such as a host's overall acceptance rate prior to their assignment to the experiment we are analyzing. As [Deng 2013] shows, appropriately chosen features, or *control variates*, can increase our statistical precision. However, having additional quasi-identifiers would likely increase information loss during the anonymization procedure. Our current implementation of generalization does not prioritize among quasi-identifiers. Thus, information essential to computing the acceptance rate gap, such as the number of accepted contacts, is as likely to be generalized as a non-essential feature, the host's prior acceptance rate. This would likely lead to information loss that would render the implementation of *control variates* unviable. Being able to prioritize among quasi-identifiers would then allow us to implement advanced methodology that increases our statistical precision.

Having a deeper understanding of data utility in our context will help us implement any of the above-mentioned approaches. It would likely be beneficial to further explore what drives the downward bias in our estimates of the acceptance rate gap for higher values of k. Some avenues for exploration include analyzing which records get suppressed or generalized when we enforce k-anonymity. Users that are more active are more likely to have their records altered (as they are more likely to have unique values for features such as number of contacts). Therefore, we can potentially increase data utility by being more careful about how we generalize such records. It may also be valuable to conduct additional simulations to understand whether k-anonymization or p-sensitization has a larger effect on data utility. This could lead us to take a different approach to enforcing k-anonymity. Revising our method of measuring the acceptance rate gap to account for the anonymity of the data may also improve data utility [Inan 2009].

## Ensuring anonymization across multiple datasets

We anticipate measuring the acceptance rate gap using this system on a regular but not continuous basis. This, in concert with our security precautions including automatic deletion of the data, has led us to work under the assumption that an attacker may gain access to one or many of the data stores/files associated with a single instance of data analysis only (one flow through the system).

A future improvement on the system may be to relax this assumption and consider each anonymized demographic dataset (**Data Store 2**) as one of many sequential or independent releases, so that an attacker may utilize multiple such datasets to attempt to achieve disclosure [Xiao 2007, Ganta 2008].

## Fully homomorphic encryption

A *homomorphic encryption scheme* is an extension of an asymmetric, or public-key, encryption scheme to allow computations on the encrypted data whose result, when decrypted, is equivalent to a result were the decrypted data computed upon directly. *Fully homomorphic encryption* (FHE) schemes allow for arbitrary computations on encrypted data, including statistical analysis on large datasets [Martins 2017, Aslett 2015, Wu 2012].

A promising area of future work is to investigate iterations on the system design using FHE to reduce risks and potentially entirely mitigate some vulnerabilities. There is also the potential to design a system whereby a separate organization provides services to allow for cryptographically secure analysis of experience gaps for technology companies such as Airbnb.

## Measuring additional experience gaps

In our current setup, we can predict the impact of anonymization on our ability to accurately measure the acceptance rate gap, or the effect of an A/B test on the acceptance rate gap. It is important to note that the impact of anonymity on data utility is relatively small in this context. This is because we have many users in each study, and many hosts and guests send or receive a small number of contacts. Therefore, we have a relatively high number of users in each equivalence class.

There are many other contexts, which are important to tackling discrimination, where there will be fewer users in each equivalence class. For example, if we were interested in examining geographical patterns of experience gaps in the United States, we would have to compute experience gaps in each geographic region, thus encountering smaller equivalence classes. There are also other experience gaps that we would like to measure. For example, we can study patterns in Instant Book cancellation rates for guests of different demographics. To ensure that the system presented in this paper can be used to measure and mitigate discrimination in such contexts, we will need to keep refining and extending it from both data utility and privacy perspectives.

Additional types of data, multiple rows of data per user, and/or high-dimensional data may bring new challenges requiring new methods, such as those articulated in [Aggarwal 2005, Fung 2010, Li 2012, Narayanan 2008, Nergiz 2008]. *Condensation* [Aggarwal 2004] is another alternative to generalization that may improve correlations across data fields for experience gap analysis. It may also require less investigation on the impact of anonymization on data utility. Related methods such as *microaggregation* [Domingo-Ferrer 2002] as well as data partitioning and data clustering algorithms mentioned in [Gkoulalas-Divanis 2014] may also prove worthwhile.

As we work to improve our measurements, develop tools, and close further experience gaps, we are committed to protecting privacy and maintaining transparency with our community.

# Conclusion

We've shown in this paper how to build a novel system to measure the acceptance rate gap while minimizing the risk that an individual's perceived race can be retrieved from the data; we achieve this by enforcing the privacy model of p-sensitive k-anonymity in order to prevent certain attribute disclosure and utilizing asymmetric encryption to ensure data flows across trust boundaries are one-way.

Our analysis of the utility of these anonymized data on the task of measuring the efficacy of interventions to reduce the acceptance rate gap shows that we can still achieve an adequate degree of statistical power to measure platform-wide changes with anonymized data.

As we have done for this initial system for measuring the acceptance rate gap, we will continue to consider the goal of privacy on equal footing to that of our ability to measure experience gaps so that we ensure privacy is properly baked into the systems we build and operate [Basu 2015, Cavoukian 2012]. We will continue to follow privacy research developments; as our system designs change, we will publish further papers for the benefit of other organizations doing similar work and to seek feedback to improve how we do our work [Sweeney 1997, Sweeney 2018].

**We know that bias, discrimination, and systemic inequities are complex and longstanding problems. Addressing them requires continued attention, adaptation, and collaboration. We encourage our peers in the technology industry to join us in this fight, and to help push us all collectively towards a world where everyone can belong.**

<pre style="white-space: pre-wrap;">

</pre>

# Appendix 1:
# Survey of values of k

The choice of an appropriate k to mitigate the risk of attribute disclosure is domain-specific and somewhat subjective. We're not aiming to thwart, for example, an individual manually reviewing data in an attempt to achieve attribute disclosure (e.g., in [Sweeney 1997] and [Sweeney 2000]). Instead, we're choosing a k so as to reasonably prevent an attack from achieving en-masse certain attribute disclosure, allowing them to target a significant number of users based upon their race—see **Technical overview: Disclosure threat categories** for further discussion of our privacy goals.

To ground our selection of k, we reviewed the literature to understand values of k used by various institutions, summarized in Table 14 below. EU and US privacy legislation doesn't provide explicit guidance around acceptable disclosure risks with respect to cell size (for the purposes of our analysis, k as described below): e.g., FERPA [Rooker 2004, Angiuli 2015], HIPAA [US DHHS 2012], EU legislation [Basu 2015, Article 29 DPWP 2014]; instead there are a variety of guidelines provided.

The operative term in many of the references in Table 14 is the *numerator* or *cell size*—the number of persons/rows/cases with a particular condition or value in a *cell*; minimum numerator before data suppression/modification is required is considered k. Some of the guidelines provide a rationale for suppression, beyond re-identification risk, based on the unreliability of statistics from small numerators, e.g., [Rhode Island DH 2016, HealthStats NSW 2015]. Many also provide additional guidelines around other aspects of the data, e.g., *denominator*—the population size, or number of persons/rows/cases aggregated in some manner with respect to the cell; what exactly constitutes the denominator can be context-specific, however [California DHS 2016].

Based on this review, we believe the following requirements for k are appropriate for measuring the acceptance rate gap:

1. Use a k value greater than or equal to 5.

2. Following [Sweeney 2000], use the highest value of k that we believe will allow us to achieve the aim of a particular analysis.

3. We may, without violating the first requirement, provide some buffer in following the second requirement in order to ensure we do not need to repeat the operation (incurring additional costs) in the event of an overly conservative (high) k.

| k | Notes |
|---|---|
| 5 | Utah Department of Health [Utah DH 2005] |
| 4 | Iowa Department of Public Health [Iowa DH 2012] |
| 10 | Washington State Department of Health [Washington DH 2018]<br><br>Previous guidelines provided k=5, since increased following 2011 adopted guidance from Center for Disease Control National Center for Health Statistics (CDC NCHS) increasing from k = 5 [Statistics 2004] to k = 10. |
| 10, 30 | Center for Disease Control National Center for Health Statistics [Klein 2002]<br><br>Survey of guidelines for data suppression for data systems for Healthy People 2010 initiative. Some of these data systems are external to CDC NCHS. [Klein 2002] |
| 3, 5 | Review of multiple guidelines for "sensitive health data" for multiple Canadian health systems. [Emam 2009] |
| 5 | Rhode Island Department of Health [Rhode Island DH 2016] |
| 5 | New South Wales Centre for Epidemiology and Evidence, following guidance from Australian Institute of Health and Welfare. [HealthStats NSW 2015] |
| 11 | California Department of Health Care Services; also cites CMS guidance. [California DHS 2016] |
| 3, 5-10 | U.S. Department of Education Privacy Technical Assistance Center, cited in [Daries 2014]; primary source no longer accessible online. |
| 5 | MIT and Harvard release of edX MOOC data set [Angiuli 2015]. |
| 5 | Patient demographic data, source cited in [Gkoulalas-Divanis 2014] |
| 5, 10, 20, 25 | Sources cited in [Sweeney 2018] for k <= 5, 10 (government policies), k <= 20 (international defamation cases), k <= 25 (US defamation). |
| 5 | Social Security Administration (SSA) [Sweeney 1997] |

**Table 14:** Surveyed values of k

# Appendix 2:
# LINDDUN privacy analysis

In this section, we provide a formal privacy analysis using the LINDDUN privacy threat modelling methodology [Deng 2011]. Threats were identified using the documented LINDDUN threat trees [DistriNet 2019]. Threat nodes from threat trees are identified in parentheses, e.g., Unawareness (U) or Re-identification possible (I_ds4).

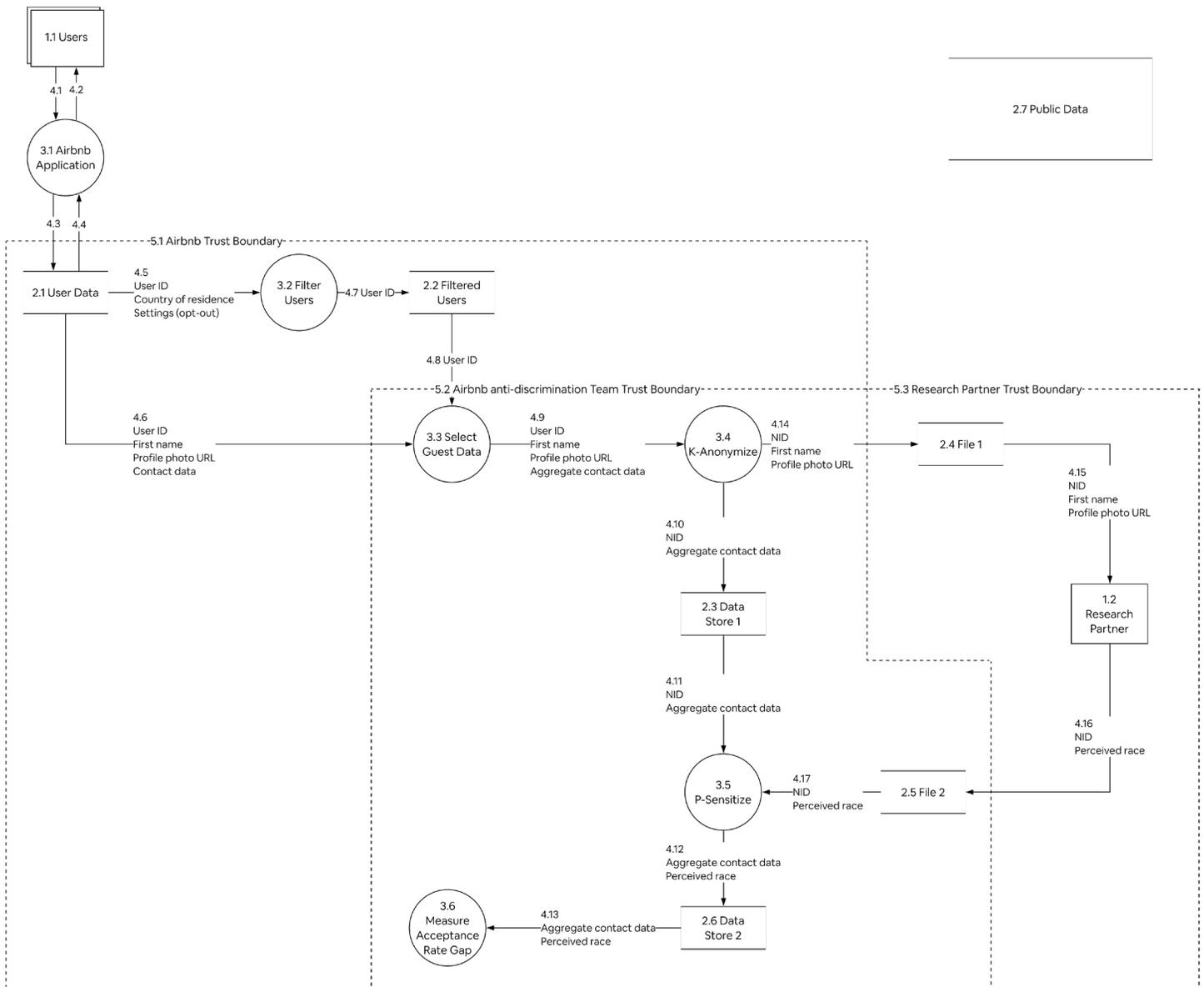

**Figure 17:** Data Flow Diagram (DFD)

# Data flow diagram

Additional details relevant to the privacy analysis (with reference to the DFD in Figure 17 on previous page):

- **2.4 File 1** is within the **5.3 Research Partner Trust Boundary** but not **5.1 Airbnb Trust Boundary** to reflect that, while those within **5.2 Airbnb anti-discrimination Team Trust Boundary** have access to **3.4 K-Anonymize** (which writes to **2.4 File 1**), the content written is asymetrically encrypted, with the public key available to those within **5.2 Airbnb anti-discrimination Team Trust Boundary** but the private key is generated and only available within **5.3 Research Partner Trust Boundary**. A similar argument applies to **2.5 File 2**'s placement.

- Data in **2.3 Data Store 1**, **2.4 File 1**, **2.6 Data Store 2**, and **2.5 File 2** are automatically deleted 30 days after persistence. This ensures data is not stored longer than necessary (`L_ds5`).

- Data in **2.4 File 1** and **2.5 File 2** are deleted by **3.5 P-Sensitize** after it completes persisting data to **2.6 Data Store 2**. This ensures data is not stored longer than necessary (`L_ds5`).

- Data in **2.3 Data Store 1** and **2.6 Data Store 2** is k-anonymous and, in the case of **2.6 Data Store 2** also p-sensitive, i.e. only as much identifying information is stored as required (`L_ds6`, `I_ds6`).

- The unique row identifiers (NID) generated by **3.4 K-Anonymize** are guaranteed to be unique per run of **3.4 K-Anonymize** only and are pseudo-random. Furthermore, we assume that no misactor can reliably reproduce the (pseudo-randomly generated) User ID ~ NID mapping, ensuring identifiability via linkability to User ID (`I_ds5`) is not achievable using NID.

- Profile photo URL (from **2.1 User Data**, **2.4 File 1**) is pseudo-randomly generated; we assume no misactor can reliably reproduce the User ID ~ Profile photo URL mapping without access to **2.1 User Data**.

# Misactor definitions

Misactors are defined based upon which trust boundaries they reside within, i.e. where they already have authorized access—the relationships between misactors and authorized trust boundaries is given in Figure 18 below. There is no misactor who holds both a role at Airbnb (**MA2**, **MA3**) and the research partner (**MA4**) so that unauthorized access is required if a misactor were to breach both **5.3 Research Partner Trust Boundary** and one of **5.1 Airbnb Trust Boundary**, **5.2 Airbnb anti-discrimination Team Trust Boundary**.

MA1. The general public, i.e. outside **5.1 Airbnb Trust Boundary** and **5.3 Research Partner Trust Boundary**. They only have authorized access to **2.7 Public Data**.

MA2. Subset of Airbnb employees within **5.1 Airbnb Trust Boundary** but not within **5.2 Airbnb anti-discrimination Team Trust Boundary**.

MA3. Subset of the Airbnb anti-discrimination team within **5.2 Airbnb anti-discrimination Team Trust Boundary**.

MA4. Subset of the research partner employees within **5.3 Research Partner Trust Boundary**.

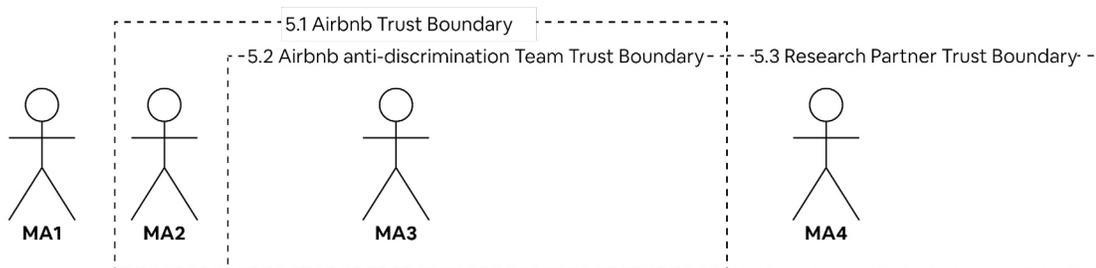

**Figure 18:** Misactor definitions

# Threat assumptions

## Threat category exclusions

TA1. Linkability threats (`L`) are not considered in isolation; instead they are only considered insofar as they contribute to other threats such as Identifiability (`I`).

TA2. Non-repudiation threats (`NR`) are not considered because plausible deniability is not a relevant privacy requirement.

TA3. Detectability threats (`D`) are not considered in isolation because internal knowledge that a user (**1.1 Users**) has participated in an analysis (**3.6 Measure acceptance rate gap**) is not considered an issue, i.e. membership disclosure threats are not considered [Gkoulalas-Divanis 2014].

TA4. Information Disclosure security threats (`ID`) are not considered in isolation; security analysis of Airbnb systems is beyond the scope of this analysis and is confidential. For the purposes of this analysis, assume that internal security best practices are followed within and across **5.1 Airbnb Trust Boundary** and **5.2 Airbnb anti-discrimination Team Trust Boundary**. An implication of this assumption is that Identifiability threats for processes (`I_p`), because they only occur in the case of Information Disclosure threats for processes (`ID_p`), are not considered in isolation.

## Component exclusions

TA5. Identifiability (`I_e`) of **1.1 Users** is not considered a threat in isolation, but only insofar as it relates to Identifiability of context data (`I_df`).

TA6. Identifiability (`I_e`) and Unawareness (`U_e`) for **1.2 Research Partner** are not considered threats.

TA7. The following are not analyzed in isolation for threats: **2.1 User Data, 3.1 Airbnb Application**, **Data flows 4.1 - 4.4**. Security/privacy analysis of these components (including e.g. authentication processes) is beyond the scope of this analysis and is confidential.

TA8. Only aspects of Non-Compliance (`NC`) pertinent to establishing the users for which analysis is permissible (**3.2 Filter Users**) are considered.

TA9. Knowledge that a user's data may be applicable for analysis is not considered sensitive, so that **2.2 Filtered Users**, **Data flows 4.5 - 4.8** are not analyzed in isolation for threats.

TA10. **Data flows 4.9 - 4.13** are not analyzed in isolation for threats; its assumed that, within **5.2 Airbnb anti-discrimination Team Trust Boundary,** Information Disclosure of a data flow (`ID_df`) requires the same or additional misactor sophistication than Information Disclosure of the relevant data store (`ID_ds`). Furthermore, security analysis of these components in Airbnb's systems is confidential.

TA11. Asymmetric encryption for **2.4 File 1** and **2.5 File 2** ensures that **Data flows 4.14 - 4.17** are sufficiently protected from Identifiability threats (`I_df4`, `I_df`).

TA12. **2.7 Public Data** is not analyzed in isolation for threats.

## Other assumptions

TA13. The data flow diagram and privacy analysis doesn't model storage or access to profile photos from the profile photo URL (**2.1 User Data**) by external entities (**1.1 Users**, **1.2 Research Partner**). Security/privacy analysis of this aspect of Airbnb's systems is confidential.

TA14. **2.4 File 1** and **2.5 File 2** are stored in an intermediate shared file service. The use of this service has been audited by the Airbnb security team and access is restricted to only the relevant subsets of the Airbnb anti-discrimination team and the research partner. We consider the intermediate shared file service outside the scope of this privacy analysis.

# Mapping to threats

An "X" in the Table 15 below represents a threat category (column) to analyze for the component (row); "X"s are grayed out if assumptions articulated above remove the necessity of analyzing the threat category for the component in the next section.

|  | L | I | NR | D | ID | U | NC |
|---|---|---|---|---|---|---|---|
| 1.1 Users | X | X |  |  |  | X |  |
| 1.2 Service Provider | X | X |  |  |  | X |  |
| 2.1 User Data | X | X | X | X | X |  | X |
| 2.2 Filtered Users | X | X | X | X | X |  | X |
| 2.3 Data Store 1 | X | X | X | X | X |  | X |
| 2.4 File 1 | X | X | X | X | X |  | X |
| 2.5 File 2 | X | X | X | X | X |  | X |
| 2.6 Data Store 2 | X | X | X | X | X |  | X |
| 2.7 Public Data | X | X | X | X | X |  | X |
| 3.1 Airbnb Application | X | X | X | X | X |  | X |
| 3.2 Filter Users | X | X | X | X | X |  | X |
| 3.3 Select Guest Data | X | X | X | X | X |  | X |
| 3.4 K-Anonymize | X | X | X | X | X |  | X |
| 3.5 P-Sensitize | X | X | X | X | X |  | X |
| 3.6 Measure Acceptance Rate Gap | X | X | X | X | X |  | X |
| 4.1 - 4.4 Data flows | X | X | X | X | X |  | X |
| 4.5 - 4.8 Data flows | X | X | X | X | X |  | X |
| 4.9 - 4.13 Data flows | X | X | X | X | X |  | X |
| 4.14 - 4.17 Data flows | X | X | X | X | X |  | X |

**Table 15:** Threat mappings

# Threat scenarios

Threat scenarios are documented as a general description of the attack in which a misactor may engage. Each scenario also includes a subjective risk assessment of high, medium, or low and an argument for the risk assessment—the likelihood of the attack scenario and its impact [Deng 2011].

### TS1: User unaware of their general participation

A user (**1.1 Users**) is unaware (`U`) of their general participation in analysis (`U_1`) to measure and mitigate discrimination by the Airbnb anti-discrimination team (**MA3**). This occurs because the notification to Airbnb community does not sufficiently raise awareness to users of their participation (`U_3`); in addition, they may not notice their option to opt out in their account's privacy & sharing page (`U_4`). Insufficient awareness may also be considered non-compliance (`NC`, `NC_4`).

We believe the likelihood of this threat is low due to our extensive work to publicize this initiative, including notifications and providing an explicit option for opt-out for US users; we believe the impact of this threat is medium as it is important that our community be aware of their participation.

### TS2: User unaware of their specific participation

A user (**1.1 Users**) is unaware (`U`) of the specific ways in which their participation in analysis by the Airbnb anti-discrimination team (**MA3**) occurs; this unawareness is due to one of:

- The general use or the timing of the use (`U_2`)—the Airbnb anti-discrimination team does not communicate each time analysis (**3.6 Measure acceptance rate gap**) occurs.

- The specific perceived race associated with their aggregate contact data (`U_5`)—because the system is designed to ensure that the perceived race data is anonymized.

We believe the likelihood of this threat is high, by design; we believe the impact of this threat is low—we do not believe a technical understanding of their individual participation in aggregate analyses is critical to our community.

## TS3: Homogeneity attack through circumventing P-Sensitize

See Figure 19 below. An authorized member of the Airbnb anti-discrimination team (**MA3**), rather than running **3.5 P-Sensitize**, directly accesses **2.5 File 2** (`I_ds1`) and, linking NID to **2.3 Data Store 1** (`I_ds4`) and then quasi-identifiers in **2.1 User Data** (`I_ds3`), achieves certain attribute disclosure (`I_ds`) for those users whose equivalence class (in **2.5 File 2**) is homogenous, i.e. has only one distinct value for perceived race. This is known as a *homogeneity attack* and may occur because **2.5 File 2** is not p-sensitive [Basu 2015, Machanavajjhala 2006].

We believe the likelihood of this threat is low as it requires the misactor be a malicious authorized member of the Airbnb anti-discrimination team (**MA3**); we believe the impact of this threat is medium—reflecting the low percentage of users we expect, based on our simulations, to be in homogenous equivalence classes (less than 1% for `k >= 5`) as shown in Figure 14 earlier in this paper.

## TS4: Homogeneity attack through unauthorized access

See Figure 20 below. An authorized Airbnb employee (**MA2**) or other misactor (**MA1**, **MA4**) gains unauthorized access to both **2.3 Data Store 1** and **2.5 File 2** (`I_ds1`) and engages in the same homogeneity attack (`I_ds`) described in **TS3** above (`I_ds3`, `I_ds4`) to achieve certain attribute disclosure (`I_ds`) only for those users whose equivalence class is homogenous.

We believe the likelihood of this threat is low as it requires a misactor to circumvent access controls to breach **5.2 Airbnb anti-discrimination Team Trust Boundary**; we believe the impact of this threat is medium as in TS3 above.

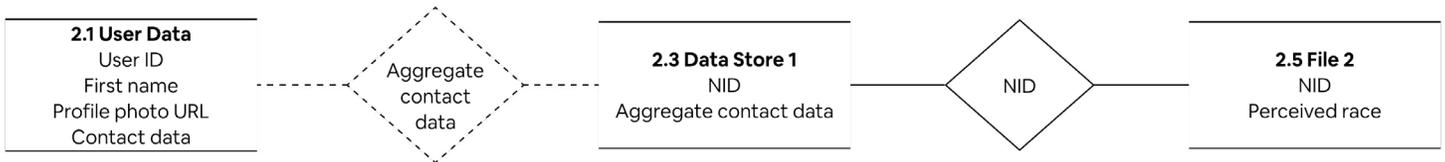

Figure 19: TS3 data-linkage diagram

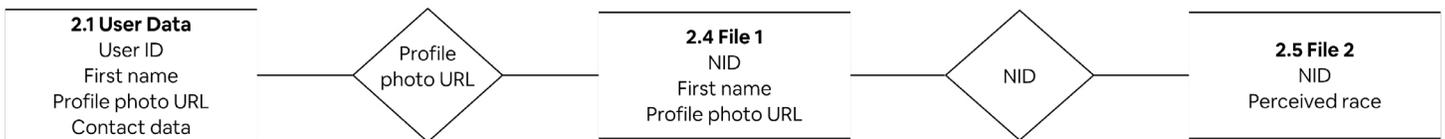

Figure 20: TS4 data-linkage diagram

## TS5: Identification through unauthorized access to research partner and circumventing P-Sensitize

See Figure 21 below. An authorized member of the Airbnb anti-discrimination team (**MA3**) gains unauthorized access to **2.4 File 1** (`I_ds1`) and, prior to running **3.5 P-Sensitize**, directly accesses **2.5 File 2** (`I_ds1`) and links NID to **2.4 File 1** (`I_ds4`) followed by the unique identifier Profile Photo to **2.1 User Data** (`I_ds3`), achieving certain attribute disclosure (`I_ds`) for all users in the analysis.

We believe the likelihood of this threat is low as it requires the misactor be a malicious authorized member of the Airbnb anti-discrimination team that also circumvents access controls to breach **5.3 Research Partner Trust Boundary**; we believe the impact of this threat is high as it leads to certain attribute disclosure for all users in the analysis.

## TS6: Identification by malicious research partner

See Figure 22 below. An authorized research partner employee (**MA4**), directly accesses **2.4 File 1, Data flow 4.16** (through circumventing the process by which **2.5 File 2** is generated; details of **1.2 Research Partner** are not diagrammed in the DFD), and **2.7 Public Data** (`I_ds1`). Facial recognition/photo matching may be used to link the user's other publicly available photos from **2.7 Public Data** to the user's Profile Photo from **2.4 File 1** (`L_ds, I_ds4`). Unless the exact same photo exists in both, the matching will be uncertain, i.e. probabilistic attribute disclosure for some users (`I_ds`). First name from **2.4 File 1** may be used to increase certainty through either its direct existence or linkage between other quasi-identifiers in **2.7 Public Data** (`I_ds6`).

We believe the likelihood of this threat is high—while we assume the likelihood of internal malicious misactors to be low, externally we assume high; we believe the impact of this threat is medium as it leads to certain or probabilistic (depending upon the photo matching method) attribute disclosure for some users in the dataset [Acquisti 2014].

## TS7: Identification through research partner unauthorized access to Airbnb

See Figure 22 below. An authorized research partner employee (**MA4**) gains unauthorized access to **2.1 User Data** (I_ds1) and accesses **2.4 File 1** and **Data flow 4.16** (see TS6 above) (`I_ds1`); they then link NID from **Data flow 4.16** to **2.4 File 1** (`I_ds4`) followed by the unique identifier Profile Photo to **2.1 User Data** (`I_ds3`), achieving certain attribute disclosure (`I_ds`) for all users in the analysis.

We believe the likelihood of this threat is low as it requires the misactor be an authorized research partner employee and breach **5.1 Airbnb Trust Boundary**; we believe the impact of this threat is high as it leads to certain attribute disclosure for all users in the analysis.

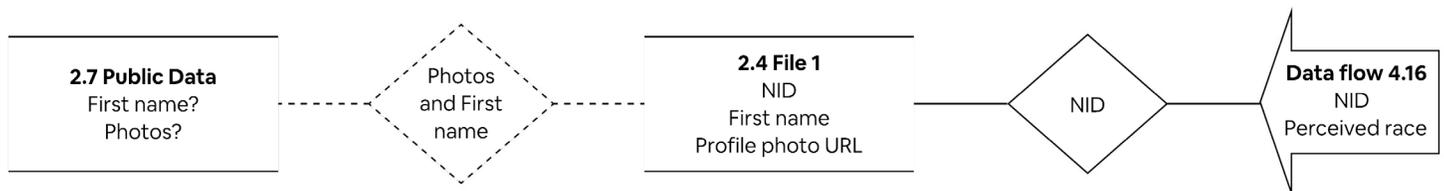

**Figure 21**: TS5 data-linkage diagram

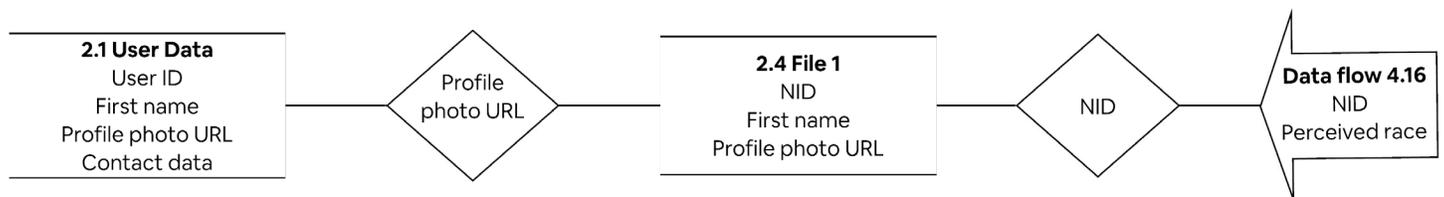

**Figure 22**: TS6 and TS7 data-linkage diagram

## TS8: Probabilistic identification by modifying posterior beliefs

See Figure 23 below. An authorized member of the Airbnb anti-discrimination team (**MA3**), linking rows in each equivalence class (>= k) in **2.6 Data Store 2** (`I_ds1`) via quasi-identifiers to rows (>= k) in **2.1 User Data** (`I_ds4`); in doing so, they may revise their posterior beliefs to have increased accuracy in estimating the perceived race of users (`I_ds`). Alternatively, an authorized Airbnb employee (**MA2**) may gain unauthorized access to **2.6 Data Store 2** (`I_ds1`) to achieve the same probabilistic attribute disclosure.

We believe the likelihood of this threat is low as it requires either (a) the misactor be a malicious authorized member of the Airbnb anti-discrimination team (**MA3**) or (b) the misactor breach **5.2 Airbnb anti-discrimination Team Trust Boundary**; we believe the impact of this threat is low given our guidelines for choosing k set forth in **Appendix 1: survey of values of k**.

## TS9: Non-compliant user participation

An authorized Airbnb employee (**MA2**) or authorized member of the Airbnb anti-discrimination team (**MA3**) modifies the policies (business logic in **3.2 Filter Users**) that ensure only users (**1.1 Users**) who have been properly notified (see TS1 above) and whose participation would not violate internal policies and any applicable laws/regulations may be considered applicable for analysis (**3.6 Measure Acceptance Rate Gap**). This modification of policies may be malicious (`NC_1`) or simply a well-intentioned mistake (`NC_2`).

We believe the likelihood of this threat is low as it requires the misactor be a malicious authorized employee or a bug in the implementation of relatively stable business logic (outside the scope of this paper); we believe the impact of this threat is high as it may lead to violation of internal policies or applicable laws/regulations.

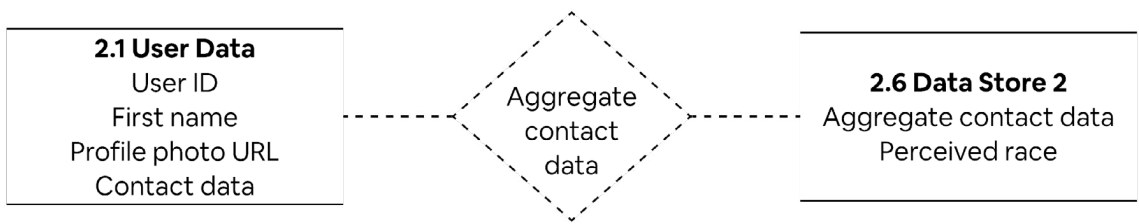

**Figure 23**: TS8 data-linkage diagram

# Appendix 3:
# Research partner requirements

This appendix is meant to precisely describe requirements and to be included in legal agreements with research partners.

## Tag Batch Process Description

### Summary

Using asymmetric encryption, Airbnb sends Research Partner a collection of photo URLs and first names that human operators, called *Taggers*, use to label them with perceived race labels, called *Tags*. The results are aggregated by Research Partner and sent to Airbnb, again using asymmetric encryption.

### Key exchange

Research Partner generates a public-private key using the GPG parameters agreed upon with Airbnb—this is the pair (Public K1, Private K1). Airbnb also generates a public-private key—this is the pair (Public K2, Private K2). Each will share only the public key (Public K1 and Public K2) with each other via a predetermined Secure Cloud Storage shared folder.

### Initiate Tag Batch

Airbnb encrypts the information for tagging in a *Tag Batch Request File* using Public K1 and uploads them to a predetermined Secure Cloud Storage shared folder. Research Partner downloads the file and decrypts them using Private K1.

### Tag Loop

For each Photo URL, First Name pair in the files, Research Partner assigns them to a Tagger who, examining the Photo and the first name, selects a perceived race label (Tag) for the most prominent person (if more than one person) in the photo, as well as the number of people in the photo (usually only one person).

### Complete Tag Batch

Research Partner collects the Tags with their associated identifier (NID) into a *Tag Batch Results File*. This file is encrypted using Public K2 and uploaded to a predetermined Secure Cloud Storage shared folder. Airbnb downloads the file and decrypts it using Private K2.

Airbnb then deletes all files in Secure Cloud Storage (Public K1 File, Public K2 File, Tag Batch Definitions File, Tag Batch Entries File, and Tag Batch Results File). Research Partner and Airbnb expunge private keys (Private K1 and Private K2) from their respective systems.

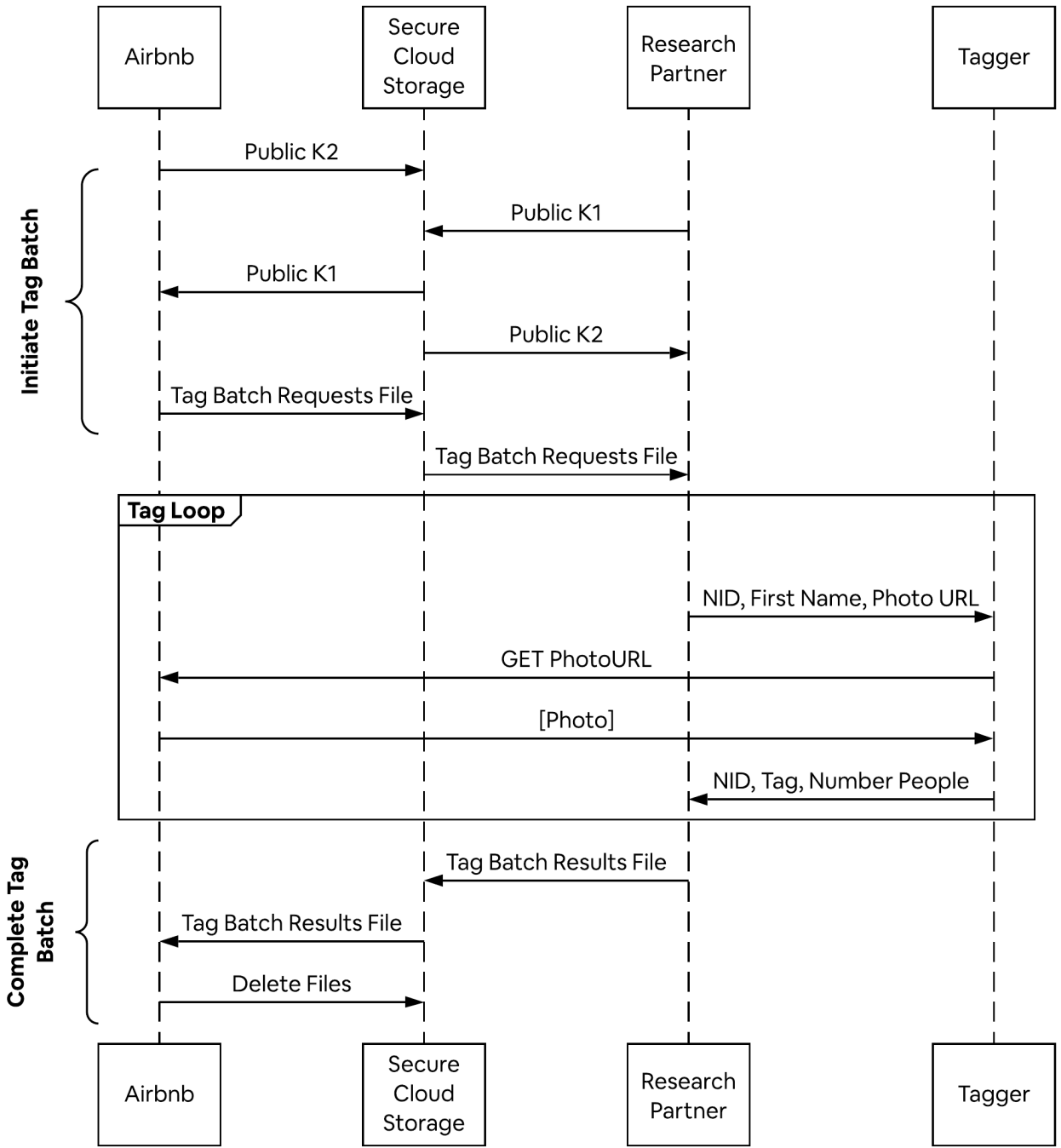

# Requirements

The key words "MUST", "MUST NOT", "REQUIRED", "SHALL", "SHALL NOT", "SHOULD", "SHOULD NOT", "RECOMMENDED", "MAY", and "OPTIONAL" in this document are to be interpreted as described in RFC 2119.

## Review Requirements

- Research Partner MUST pass Airbnb Research Partner Review, including an Airbnb Security Review.
- Research Partner MUST pass an Airbnb Security Review at-least once every 12 months.
- Research Partner SHALL notify Airbnb of changes to Research Partner systems they believe, based on previous Airbnb Security Reviews, SHOULD be relevant; SHOULD Airbnb deem it necessary, Research Partner MUST pass an Airbnb Security Review.

## Data / Information Requirements

- Research Partner MUST delete data associated with a Tag Batch from internal systems within 30 days of transmitting Tag Batch Results File to the predetermined Secure Cloud Storage shared folder.
- Research Partner MUST NOT share Private K1 with Airbnb or any other party.
- Research Partner SHALL notify Airbnb if they suspect any of the following MAY have been compromised or otherwise disclosed: Public K1, Public K2, Private K1, Private K2.
- Research Partner MUST NOT utilize any information exchanged and derived from the process described above, including Tags, for any use beyond that described in this requirements document.
- Research Partner MUST NOT reveal the identity of Airbnb to Taggers or to any third parties (confidentiality).
- The parameters used for GPG keys SHALL be:

```
Key-Type: RSA
Key-Length: 2048
Name-Real: <Research Partner name>
Name-Email: <Research Partner email address>
Expire-Date: seconds=5256000
Passphrase: <Chosen by Research Partner for each Tag Batch>
```

## Process Requirements

- Research Partner Taggers SHALL be humans, i.e. Research Partner MUST NOT utilize computer algorithms (including machine learning models) to select Tags in place of or in addition to human decision-making.
- Research Partner Taggers SHALL be based in the United States.
- Research Partner Taggers SHALL select Tags based solely upon a subjective visual review of the photos and associated first names. Research Partner and Research Partner Taggers MUST NOT utilize objective measurement tools such as rulers in place of or in addition to subjective visual review.

- Research Partner MUST assign a unique identifier to each Tagger that is used to uniquely identify that Tagger for the entirety of a single Tag Batch.
- Tag Batch Results File sent to Airbnb by Research Partner:
  - MUST contain the associated ID (NID) from the Tag Batch Request File.
  - MUST contain the unique identifier (TID) assigned to the Tagger by Research Partner.
  - MUST contain a Tag representing the perceived race of the most prominent person in the photo from the photo URL that is one of the allowed labels articulated in the current SoW.
  - MUST contain the number of persons in the photo from the photo URL. If no person is contained in the photo, this value SHALL be 0.

# Informal File Specifications

### Tag Batch Requests File

```
<NID_1>, <PublicPhotoURL_1>, <FirstName_1>
[..]
<NID_N>, <PublicPhotoURL_N>, <FirstName_N>
```

### Tag Batch Results File

```
<NID_1>, <TID_1>, <Tag_1>, <NumPeopleInPhoto_1>
[..]
<NID_N>, <TID_N>, <Tag_N>, <NumPeopleInPhoto_N>
```